\documentclass[twocolumn]{aastex61}

\newcommand\aastex{AAS\TeX}

\newcommand{\wt}{\color{white}}

\newcommand{\ie}{{\it i.e.}}
\newcommand{\eg}{{\it e.g.}}
\newcommand{\goes}{{\it GOES}}
\newcommand{\sdo}{{\it SDO}}
\newcommand{\rhessi}{{\it RHESSI}}
\newcommand{\asecs}{\mbox{\ensuremath{^{\prime\prime}}}}
\received{\today}
\revised{\today}
\accepted{\today}
\submitjournal{ApJ}
\shorttitle{\aastex\ sample article}
\shortauthors{Hernandez-Perez et al.}

\begin{document}

\title{AASTEX: Generation mechanisms of quasi-parallel and quasi-circular flare ribbons in a confined flare}

\correspondingauthor{Aaron Hernandez-Perez}
\email{aaron.hernandez-perez@uni-graz.at}

\author{Aaron Hernandez-Perez}
\affil{IGAM/Institute of Physics, University of Graz, A-8010 Graz, Austria}

\author{Julia K. Thalmann}
\affil{IGAM/Institute of Physics, University of Graz, A-8010 Graz, Austria}

\author{Astrid M. Veronig}
\affil{IGAM/Institute of Physics, University of Graz, A-8010 Graz, Austria}

\author{Yang Su}
\affil{Key Laboratory of Dark Matter \& Space Astronomy, Purple Mountain Observatory\\
Chinese Academy of Sciences, 2 West Beijing Road, 210008 Nanjing, China}

\author{Peter G\"om\"ory}
\affil{Astronomical Institute, Slovak Academy of Sciences, 05960 Tatransk\'a Lomnica, Slovakia}

\author{Ewan C. Dickson}
\affil{IGAM/Institute of Physics, University of Graz, A-8010 Graz, Austria}

\begin{abstract}
We analyze a confined multiple-ribbon M2.1 flare (SOL2015-01-29T11:42) that originated from a fan-spine coronal magnetic field configuration, within active region NOAA 12268. The observed ribbons form in two steps. First, two {\it primary} ribbons form at the main flare site, followed by the formation of {\it secondary} ribbons at remote locations. We observe a number of plasma flows at extreme-ultraviolet temperatures during the early phase of the flare (as early as 15 min before the onset) propagating towards the formation site of the secondary ribbons. The secondary ribbon formation is co-temporal with the arrival of the pre-flare generated plasma flows. The primary ribbons are co-spatial with \rhessi\ hard X-ray sources, whereas no enhanced X-ray emission is detected at the secondary ribbons sites. The (E)UV emission, associated with the secondary ribbons, peaks $\sim$1~min after the last \rhessi\ hard X-ray enhancement. A nonlinear force-free model of the coronal magnetic field reveals that the secondary flare ribbons are not directly connected to the primary ribbons, but to regions nearby. Detailed analysis suggests that the secondary brightenings are produced due to dissipation of kinetic energy of the plasma flows (heating due to compression), and not due to non-thermal particles accelerated by magnetic reconnection, as is the case for the primary ribbons.

\end{abstract}

\keywords{keyword -- keyword -- keyword -- keyword}

\section{Introduction}
     \label{S-Introduction} 

Solar flares are among the most energetic phenomena in the solar corona, and are frequently associated with coronal mass ejections \citep[][]{2006ApJ...650L.143Y,2008ApJ...673.1174Y}, which have the most significant influence on our space weather conditions on Earth \citep[\eg,][]{1991JGR....96.7831G}. In one interpretation, the ``standard" model of eruptive flares \citep[based on the work of][]{1964NASSP..50..451C,1966Natur.211..697S,1974SoPh...34..323H,1976SoPh...50...85K}, they are driven by an erupting filament that stretches the embedding magnetic field when moving upwards. In its wake, oppositely directed magnetic fields are drawn towards each other, to form a current sheet in which magnetic reconnection sets in and releases large amounts of energy \citep[\eg,][]{2000JGR...10523153F}. The process of reconnection is also accompanied by plasma heating, bulk flows and particle acceleration, up to relativistic energies \citep[for reviews see, \eg,][]{2002A&ARv..10..313P,2011LRSP....8....6S}. Electrons, accelerated to non-thermal energies in and around the reconnection region, spiral along the newly reconnected magnetic field towards the denser lower solar atmosphere, producing X-ray emission. The resulting signatures are often observed in the form of flare kernels or ribbons \citep[for reviews see][]{2011SSRv..159...19F,JGRA:JGRA53082,2017LRSP...14....2B}.

Despite its relative success, the ability of the purely two-dimensional standard model for {\it eruptive} (\ie, CME-associated) flares is limited as flares are an intrinsically three-dimensional process. Therefore, the standard model has been extended to three-dimensions recently \citep[for a recent review see][]{2015SoPh..290.3425J}.  Typically observed flare-associated features, that cannot be explained by the standard 2D model are, the expansion of flare ribbons along the polarity inversion line (PIL) prior to the separating ribbon motion \citep[\eg,][]{2004SoPh..222..279F,2012ApJ...744...48C}, as well as the observed shape of the flare ribbons deviating from the ``classical'' quasi-parallel appearance, in the form of two ribbons. Such deviations may appear in the form of multiple quasi-parallel ribbons \citep[\eg,][]{2014ApJ...781L..23W,2016ApJ...829L...1L}, $J$-shaped ribbons \citep[\eg,][]{2009SoPh..258...53C,2014ApJ...788...60J}, and circular ribbons. The last are of particular interest, as they are thought to be associated with a coronal fan-spine topology \citep[][]{1990ApJ...350..672L}, characteristic for {\it confined} (\ie, CME-less) flares. Flare-associated ribbon emission characteristic for such a coronal topology includes quasi-circular ribbons, resembling the footprint of the coronal fan-dome with the low atmosphere, a more compact central and a possibly elongated remote ribbon. The compact and elongated ribbons mark the intersections of the inner and outer spine field lines and the photosphere, respectively \citep[see, \eg,][]{2009ApJ...700..559M,2012A&A...547A..52R,2014ApJ...792...40V}. Occasionally observed are a varying combination of plasma flows, along the inner  \citep[][]{2017arXiv170300665R} and outer spine and fan field \citep[][]{2011ApJ...728..103L}, as well as X-ray jets \citep[][]{2010ApJ...714.1762P,2012ApJ...760..101W}. 

Studies were presented on flares exhibiting complex patterns of flare ribbons, including quasi-parallel, -circular, as well as a central and a remote ribbon \citep[\eg,][]{2013ApJ...778..139S,2015ApJ...812L..19L,2016A&A...591A.141J, 2016ApJ...832...65Z}. Others reported similar cases but lacking observations of a central and/or remote brightening \citep[\eg,][]{2009ApJ...704..341S,2015ApJ...812...50J,2012ApJ...760..101W}. All of the aforementioned studies suggest the simultaneous presence of at least two distinct flux systems in the corona: a flux rope (its instability causing the subsequent parallel ribbons) and a coronal fan dome (its footprint in the lower atmosphere observed in the form of a circular ribbon). Furthermore, the analyzed events involve the destabilization of the underlying flux-rope system and subsequent reconnection at a coronal null point, which lead to a breakout-type reconnection producing a subsequent CME, or to X-ray jet activity \citep[in the case of][]{2012ApJ...760..101W}.

In order to address the original cause of the flare ribbon emission, the relative timing of flare-associated emission can be analyzed, most importantly at X-ray and (E)UV wavelengths, in context with the associated coronal magnetic field. Magnetic field previously involved in magnetic reconnection (as traced from flare kernels) allows us to understand the spatial organization of the observed emission. This includes, \eg, that the temporal and spatial evolution of strongest (E)UV emission is well correlated with that of thermal X-ray sources during flares. Such emission stems from heated coronal flare plasma and is observed to connect ribbons seen in UV, implying a newly established magnetic connection \citep[see, \eg,][for a recent study]{2016ApJ...826..143T}. Non-thermal hard X-ray (HXR) emission is most often found in the form of one or two compact sources on either side of a PIL \citep[\eg,][]{2001SoPh..204...69F,2012ApJ...746...17G,2007CEAB...31...49T}, apart from rare observations of entire ribbons \citep[\eg,][]{2007ApJ...658L.127L}. They are thought to result from thick-target bremsstrahlung of the high-energy electrons that penetrate the lower, thus denser, atmospheric layers \citep[see, \eg, reviews by][]{1988SoPh..118...49D,2011SSRv..159..107H}. The observed flare kernel/ribbon emission can in some events also be explained by alternative scenarios, however, as discussed in the following. 

\cite{2014ApJ...782L..27Z} for instance, separated flare ribbons (FRs) into two groups: {\it normal} FRs (NFRs), connected by post-flare loops, and {\it secondary} flare ribbons (SFRs), which are not connected by post-flare loops. SFRs are observed to occur in events with complicated magnetic topologies and their generation mechanism is still unclear. In their study, 19 X-class flares were investigated. The SFRs were separated into two further groups depending on their appearance in time relative to the NFRs, namely, simultaneous or delayed. It was speculated that the formation of the SFRs appearing simultaneously with the NFRs was due to a disturbance resulting from the main flare. As for the delayed SFRs, it was speculated that they may be formed due to large-scale secondary magnetic reconnection triggered by the initial magnetic reconnection at the main flare site \citep[][]{ 2011ApJ...739...59W, 2013ApJ...778L..36L, 2013ApJ...778..139S}. 

Another possibility for the generation of chromospheric brightenings was reported by \cite{2013ApJ...776L..12G}, in which the non-escaping material from a filament eruption fell downwards under the influence of gravity. The falling plasma compressed the chromosphere, dissipating the kinetic energy and producing EUV brightenings due to energy release at the impact locations. In that case, the locations of the brightenings were not connected by post-flare loops, and therefore presenting a possible physical mechanism for SFR formation.

\cite{2009ApJ...700..559M} analyzed a C-class flare that originated from a coronal fan-spine topology that divided the AR into two connectivity domains, each of them including a spine separatrix field. The flare exhibited a quasi-circular ribbon, which was associated with the fan separatrix surface that originated from the null point, and its generation was attributed to accelerated particles from the reconnection site at a null point. Two other (elongated) ribbons (SFRs) were observed, one of them associated with the inner spine co-temporal with the quasi-circular ribbon, and the other at a remote location, associated with the outer spine, forming $\sim$30s later. It was concluded that these SFRs could not be directly linked to thick-target bremsstrahlung and their generation was attributed to the presence of quasi-separatrix layers \citep{1996JGR...101.7631D} surrounding the spine field lines. \cite{2012A&A...547A..52R} subsequently analyzed the relationship between X-ray and UV emission for the same event, and found a direct correlation between the time at which the SFR at the end of the outer spine was formed and an enhancement in the \rhessi\ flux for the 25--50 keV energy range. They suggested that, as reconnection proceeds, the magnetic field lines, which undergo slip-running reconnection \citep{2006SoPh..238..347A}, would ``slip'' towards the null point. Hence, a higher flux of accelerated particles would be able to flow along the outer spine subsequently, producing the SFR. The time delay of the formation of the SFR could be attributed to the time of the slipping of the magnetic field in involving the fields linked to that location in the chromosphere.

To our knowledge, no confined flare has been reported so far exhibiting all three kinds of ribbon signatures (quasi-parallel, -circular, remote) posing challenges to traditional fan-spine (null-point) related reconnection scenarios, as the emission that originated from the {\it secondary} ribbons exhibits peculiar spatial and temporal behavior. Based on this, we propose an alternative explanation under which circumstances {\it secondary}/remote brightenings may be caused during confined solar flares. 

\section{Data and Modeling}
     \label{S-Data}

\subsection{Observational data}

The Atmospheric Imaging Assembly \citep[AIA;][]{2012SoPh..275...17L} on board the {\it Solar Dynamics Observatory} \citep[\sdo;][]{2012SoPh..275....3P}, provides high-resolution full-Sun images. It consists of four telescopes, optimized to observe UV and EUV emission from the solar atmosphere, with a spatial resolution of $1\farcs5$. The chromosphere, transition region and the quiet corona are studied using AIA 304 and 171~\AA\ filtergrams (temperature response peaking at about 50000\,K and 0.6\,MK, respectively). The active-region corona is monitored using AIA 211~\AA\ images, sensing plasma at temperatures of $\sim$2\,MK. Hot, flare plasma is studied using AIA 94~\AA\ ($\sim$6.3\,MK), and AIA 131~\AA\ images, which samples coronal plasma with temperatures around 0.4 and 10\,MK. All images were co-registered, co-aligned, and differentially de-rotated with respect to the time of the \goes\ 1--8~\AA\ peak (at 11:42~UT), using standard IDL {\it SolarSoft} procedures.

For visualization purposes, all AIA observations presented in this paper were processed by the noise adaptive fuzzy equalization method \citep[NAFE;][]{2013ApJS..207...25D} to enhance visibility of detected fine structures. NAFE is an image processing method that improves visualization of fine structures in AIA images. It provides intensity enhanced images of better quality. The main parameters of the code are $\gamma$ and $w$. $\gamma$ influences the brightness of the final (processed) image. Higher values of $\gamma$ lead to brighter processed images. The constant $w$ is the so-called NAFE weight and characterizes the level of image enhancement. The value of the NAFE weight is enclosed within the interval (0,0.3), where a zero value gives images without any enhancement and a value of 0.3 corresponds to extreme intensity enhancement in the processed images. The detailed mathematical explanation and meaning of these parameters is given in \cite{2013ApJS..207...25D}. For all EUV channels, we chose $\gamma=2.6$ and $w=0.25$. For UV channels, $\gamma=2.2$ and $w=0.2$. Since the unprocessed filtergrams for individual channels have different dynamic ranges, the scaling parameters that determine the minimum and maximum values of input and output images were set differently for all AIA channels. However, they were kept constant for the whole data set at each particular wavelength. These NAFE parameters were determined by visual inspection for the processed images in order to deliver the best results. 

The magnetic characteristics of the AR under study are determined from photospheric line-of-sight (LOS) magnetic field data, based on polarization measurements from the \sdo\ Helioseismic and Magnetic Imager \citep[HMI;][]{2012SoPh..275..229S}. It performs full-disk measurements in the Fe~{\sc I} 6173~\AA\ line with a spatial resolution of $\sim1''$.

X-ray images for the studied AR were reconstructed from observations from the {\it Ramaty High Energy Spectroscopic Imager} \citep[\rhessi;][]{2002SoPh..210....3L}. \rhessi\ observes X-rays/gamma-rays at energies above 3~keV with a time cadence of 4~seconds, a spatial resolution of $2\asecs$ at X-ray energies up to $\sim$100~keV, $7\asecs$ for X-ray energies up to $\sim$400~keV and spectral resolution of $\sim$1~keV. In order to produce X-ray images, \rhessi\ data from the front detectors 3--9 were supplied to the CLEAN algorithm \citep{2002SoPh..210...61H}. X-ray spectra were obtained using data from the front segment of detector 5, at energies in the range $\sim$4--300~keV, as its spectrum most closely resembles the mean spectrum based on the measurements of all detectors. For the spectral fitting we use an isothermal model and a thick-target non-thermal emission model \citep{2008A&A...486.1023B,2003ApJ...595L..97H}.

To estimate the temperature and density of the observed plasma structures the Sparse Differential Emission Measure (DEM) inversion code \citep{2015ApJ...807..143C} was used to calculate the total Emission Measure.

\subsection{Magnetic field modeling}
     \label{S-Magnetic field modeling}

The 3D coronal magnetic field configuration in and around NOAA~12268 was modeled based on full-disk vector magnetic field observations from HMI \citep{2014SoPh..289.3483H}, in particular the {\sf hmi.B\_720s} data series which provides the total field, inclination and azimuth on the entire solar disk. After disambiguation of the provided azimuth\footnote{For details of the procedure see: \url{http://jsoc.stanford.edu/jsocwiki/FullDiskDisamb}},
the image-plane magnetic field vector is derived and de-projected in order to obtain the true (local) field \citep{1990SoPh..126...21G}. A sub-field, covering the flaring AR as well as its quiet-Sun surroundings, was used as an input to a nonlinear force-free (NLFF) model scheme \citep[for details see][and Sect.\ 2.2.1 of \citeauthor{2015ApJ...811..107D}~\citeyear{2015ApJ...811..107D}]{2010A&A...516A.107W}.

The native (full-resolution) pixel scale of the photospheric field data is about 360~km pixel$^{-1}$ (0\farcs504). For NLFF modeling, we binned the data to 720~km\,pixel$^{-1}$ (about 1\farcs01) and adopted a computation domain of $242.5 \times 219.4 \times 121.3$\,Mm$^3$. The vertical flux at the lower boundary of the computational domain is balanced to within about 5\% and the binning of the data is nearly flux preserving ($\Delta \phi \approx 2\%$). The method of \cite{2010A&A...516A.107W} modifies the ``original'' (input) data twice, once during preprocessing \citep[which finds force-free consistent boundary data from the observed data; for details see][]{2006SoPh..233..215W} and once during the NLFF reconstruction itself. The corresponding changes to the measured (input) vertical flux amount to $\Delta \phi \approx 3\%$. Thus, changes to the vertical flux due to binning of the data and NLFF modeling are on the order of the periodic variations due to the orbital motion of \textit{SDO} \citep[][]{2012SoPh..279..295L}. The corresponding changes to the horizontal magnetic field, $B_{\rm h}$, amount to $\Delta \langle B_{\rm h} \rangle \approx [18,32]\%$ from binning/NLFF modeling, respectively. Importantly, these changes are most pronounced in weak-field regions \citep[see][for details]{2015ApJ...811..107D}. 

Finally, we list two controlling parameters in order to quantify the goodness of the obtained NLFF coronal magnetic field solution: (1) For the current-weighted average of the sine of the angle between the modeled magnetic field and electric current density we find $CWsin \simeq 0.1$. (2) For the volume-averaged fractional flux we find $\langle |f_i| \rangle \simeq 10^{-4}$. Note that for a perfectly force-free and solenoidal solution, $CWsin=0$ and $\langle|f_i|\rangle=0$ \citep[for details see, \eg,][]{2000ApJ...540.1150W,2006SoPh..235..161S}.

\section{Results}
     \label{S-Results}

\subsection{Event Overview}
     \label{S-Event Overview}

Active region (AR) NOAA~12268 emerged on the east limb on 2015 January 21 and rotated over the west limb on February 4. During disk passage, it was very prolific in producing confined flares (6 flares of \goes\ class M1.0 or larger). In this study, we concentrate on the M2.1 (1B) flare on January 29, peaking at 11:42~UT (SOL2015-01-29T11:42M2.1). 

%%%%%%%%%%%%%%%%%%%%%%%%%%%%%%%%%%%% FIGURE 1 %%%%%%%%%%%%%%%%%%%%%%%%%%%%%%%%%%%%
\begin{figure*}[ht]
\centerline{
\centering\includegraphics[width=\textwidth]{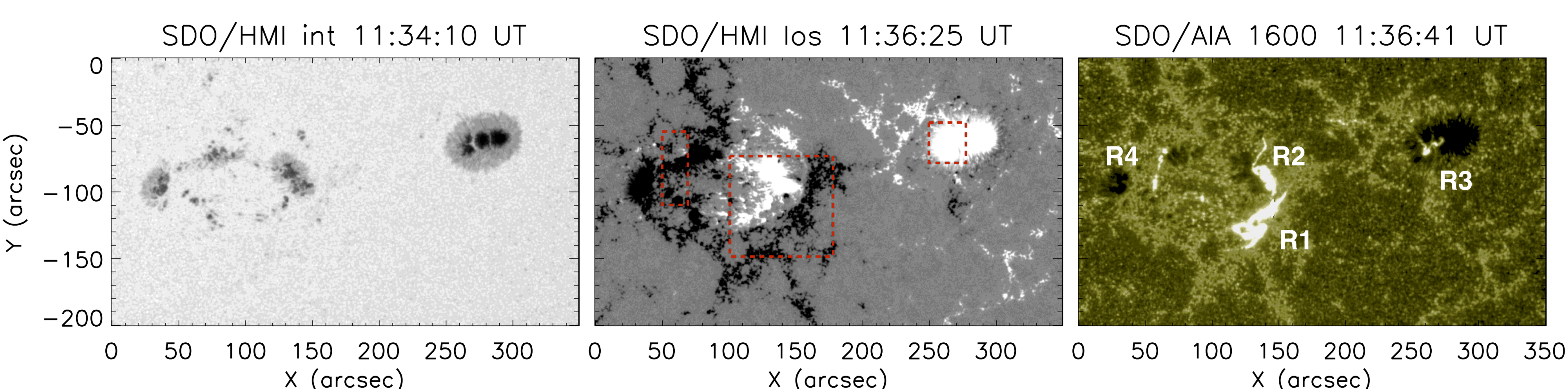}
\put(-473,100){\sf\color{black}(a)}
\put(-315,100){\sf\color{white}(b)}
\put(-157,100){\sf\color{white}(c)}
}
\caption{
AR~12268, as observed on 2015 Jan 29, during the impulsive phase of an M2.1 flare. (a) HMI continuum emission. (b) HMI LOS magnetic field, scaled to $\pm$400~G. Black/white color indicates negative/positive polarity. The red dashed rectangles mark selected regions discussed in the text for a brightness analysis in Sec.~\ref{S-Spatial and temporal correspondences of UV and HXR emission}. (c) AIA 1600~\AA\ emission at the time of the HXR peak. The ribbons are named R1, R2, R3 and R4 for later reference.
}
\label{continuum_LOS_AIA1600_as}
\end{figure*}
%%%%%%%%%%%%%%%%%%%%%%%%%%%%%%%%%%%%%%%%%%%%%%%%%%%%%%%%%%%%%%%%%%%%%%%%%%%%%%%%%%%

Fig.~\ref{continuum_LOS_AIA1600_as} summarizes the main features of AR~12268 during the impulsive phase of the flare. The negative polarity of the LOS magnetic field (Fig.~\ref{continuum_LOS_AIA1600_as}b) exhibits a horseshoe-like shape, encompassing a region of positive polarity (parasitic polarity). The main (leading) sunspot on the west of the AR, also has positive polarity (compare Fig.~\ref{continuum_LOS_AIA1600_as}a and ~\ref{continuum_LOS_AIA1600_as}b). Multiple flare ribbons are visible in different parts of the AR at that time (Fig.~\ref{continuum_LOS_AIA1600_as}c). One of the ribbons is located within the western (leading) positive-polarity sunspot (located around $(x,y)=(270\asecs,-70\asecs)$ and labeled ``R3''). Three ribbons appear in the eastern part of the AR. One of them sits within the ``parasitic'' positive-polarity region (located around $(x,y)=(140\asecs,-80\asecs)$ and and labeled ``R2''), and two in the negative polarity region that encompasses the parasitic polarity (marked as ``R1'' and ``R4''). R1 and R2 form a pair of elongated primary ribbons and exhibit the strongest observed UV emission and are largest in extent. The secondary ribbons, R3 and R4, are much smaller in extent and less intense. In the top panel of Figure~\ref{SDO_AIA_1600_SDO_AIA_304_LINEAR}, we show the integrated \goes\ SXR flux and \rhessi\ X-ray count rates, for the pre-flare, impulsive (indicated by the shaded area) and decay phase of the flare. Note that the nominal end time of the flare was around 11:52~UT, but \rhessi\ entered night time already at 11:50~UT. The SXR flux starts to increase at 11:34~UT (start of the impulsive phase) and peaks at 11:38~UT (end of the impulsive phase) for the \rhessi\ 3--6, 6--12 and 12--25 keV count rates and at 11:42~UT for \goes\ 1--8~\AA. Three distinct peaks in the non-thermal emission (25--50~keV) are observed, around 11:35:50~UT, 11:36:10~UT and 11:37:10~UT. During the decay phase, fluctuations in the 6--12~keV and 12--25~keV energy bands (dominated by thermal emission) are clearly discernible.

\subsection{(E)UV and X-ray flare morphology}
     \label{S-(E)UV and X-ray flare morphology}

%%%%%%%%%%%%%%%%%%%%%%%%%%%%%%%%%%%% FIGURE 2 %%%%%%%%%%%%%%%%%%%%%%%%%%%%%%%%%%%%

\begin{figure*}[ht]
\centerline{
\includegraphics[width=\textwidth]{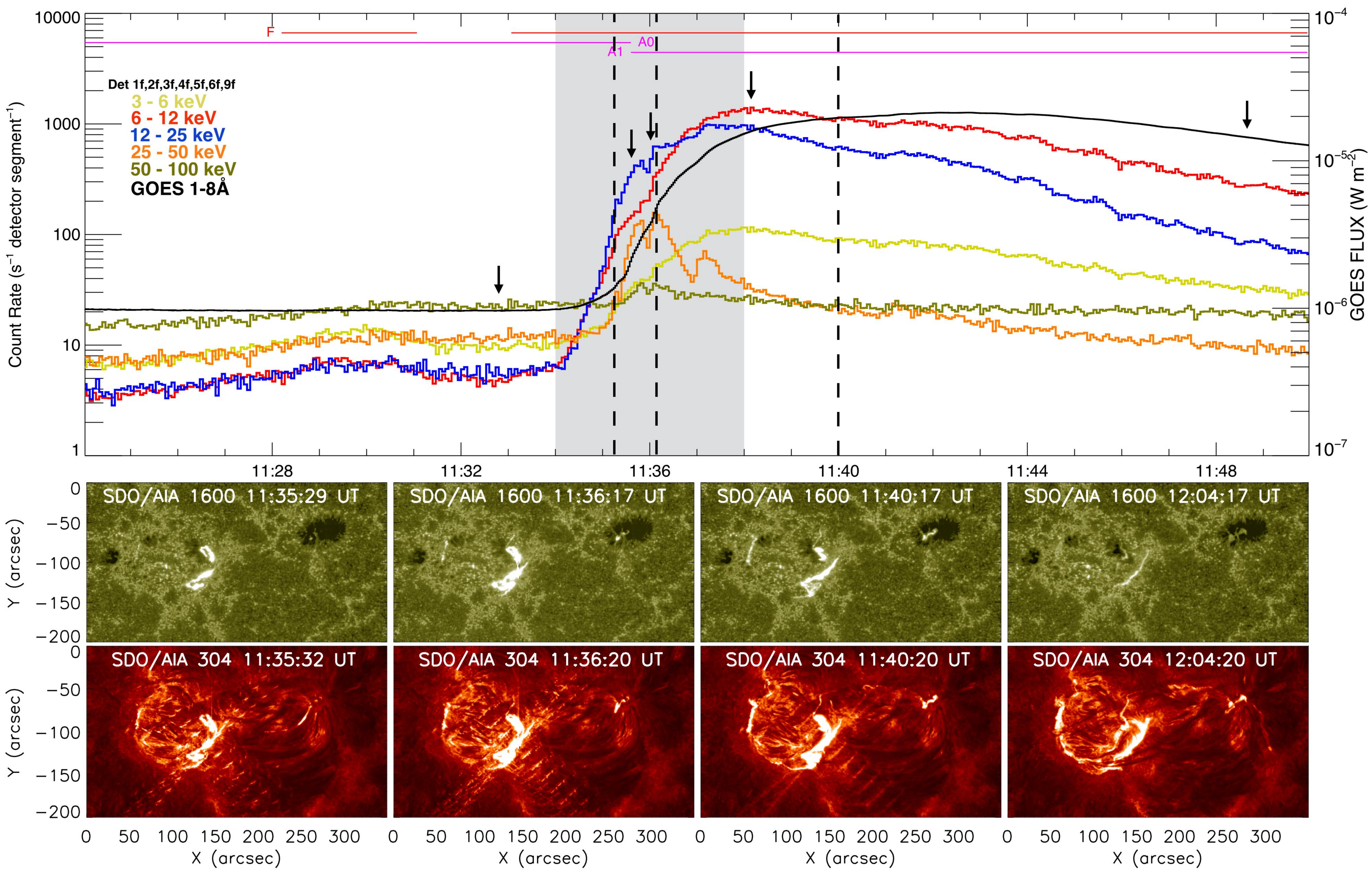}         }
\caption{
Top: \rhessi\ X-ray count rates and \goes\ SXR flux during the M2.1 flare. \rhessi\ lightcurves were constructed using the front segments of detectors 1, 3, 4, 5, 8 and 9. The shaded area indicates the impulsive phase of the flare. Bottom: Time evolution of the flare-associated emission in the chromosphere (AIA 1600~\AA\ top panels) and chromosphere/transition region (AIA 304~\AA\ bottom panels), in the course of the M2.1 flare. The first two panels show the (E)UV flare ribbon emission in the early and mid impulsive phase and the third and fourth panels the early and late decay phase (indicated by black dashed lines in the lightcurve except for the fourth panel because \rhessi\ entered night time at 11:50~UT). The black arrows indicate the times at which the \rhessi\ spectra was performed in Fig.~\ref{rhessi}. In the Electronic Supplementary Material, a movie is attached to this figure. Movie 1 shows the evolution of the flare in co-temporal 1600 and 304 \AA \ maps.}
\label{SDO_AIA_1600_SDO_AIA_304_LINEAR}
\end{figure*}  
%%%%%%%%%%%%%%%%%%%%%%%%%%%%%%%%%%%%%%%%%%%%%%%%%%%%%%%%%%%%%%%%%%%%%%%%%%%%%%%%%%%

The middle and bottom panels of Fig.~\ref{SDO_AIA_1600_SDO_AIA_304_LINEAR} display a time sequence of the flare associated chromospheric and chromospheric/transition region emission, in 1600~\AA\ (sensing plasma at temperatures of $\sim$6000\,K and $\sim$100000\,K) and 304~\AA\ (sensing plasma at temperatures of $\sim$50000\,K) respectively. The first two panels show the (E)UV flare ribbon emission in the early and mid impulsive phase and the third and fourth panels the early and late decay phase (indicated by black dashed lines in the lightcurve except for the fourth panel because \rhessi\ entered night time at 11:50~UT). During the early impulsive phase, at around 11:34:47~UT, two primary ribbons (labeled as ``R1'' and ``R2'' in Fig.~\ref{continuum_LOS_AIA1600_as}c) start to form. This time corresponds to the increase in the \rhessi\ SXR (6--25~keV), and they are fully developed at around 11:35:30~UT, which corresponds to the sharp increase in the \goes\ SXR flux and \rhessi\ 6--25~keV and 25--50~keV count rates. At 11:36:20~UT the primary ribbons are most prominent, tightly related in time with the strongest HXR (25--50~keV) peak (which will be discussed in Fig.~\ref{lightcurves}). Shortly prior to the peak UV emission of the primary ribbons, the secondary ribbons (R3 and R4) start to form and are fully developed around 11:40:20~UT. Comparison of the flare-ribbon emission observed in UV (upper panels of the image sequence in Fig.~\ref{SDO_AIA_1600_SDO_AIA_304_LINEAR}) and EUV (lower panels of the image sequence in Fig.~\ref{SDO_AIA_1600_SDO_AIA_304_LINEAR}) shows that the primary ribbons (\ie, R1 and R2) evolve in the form of quasi-parallel ribbons (see AIA 304 \AA\ image at 12:04~UT in Fig.~\ref{SDO_AIA_1600_SDO_AIA_304_LINEAR}) and that R1 and R4 actually mark segments of an extended quasi-circular flare ribbon. Particularly, as observed in AIA 304 \AA, the quasi-circular ribbon does not brighten in a sequential manner. R1 forms first, R4 forms later, and only then the space in-between these two ribbons (\ie, connecting them) fills out and becomes more prominent (see the rightmost lower panel of the image sequence in Fig.~\ref{SDO_AIA_1600_SDO_AIA_304_LINEAR}). In addition, a further ribbon-like structure appears at $x=[185\asecs$--$240\asecs]$, $y=[-50\asecs]$ aligned with the E-W direction.  In the Electronic Supplementary Material, a movie is attached to Fig.~\ref{SDO_AIA_1600_SDO_AIA_304_LINEAR}. Movie 1 shows the evolution of the flare in AIA 1600 and 304 \AA.

%%%%%%%%%%%%%%%%%%%%%%%%%%%%%%%%%%%% FIGURE 3 %%%%%%%%%%%%%%%%%%%%%%%%%%%%%%%%%%%%

\begin{figure*}[ht]
\centerline{
\includegraphics[width=\textwidth]{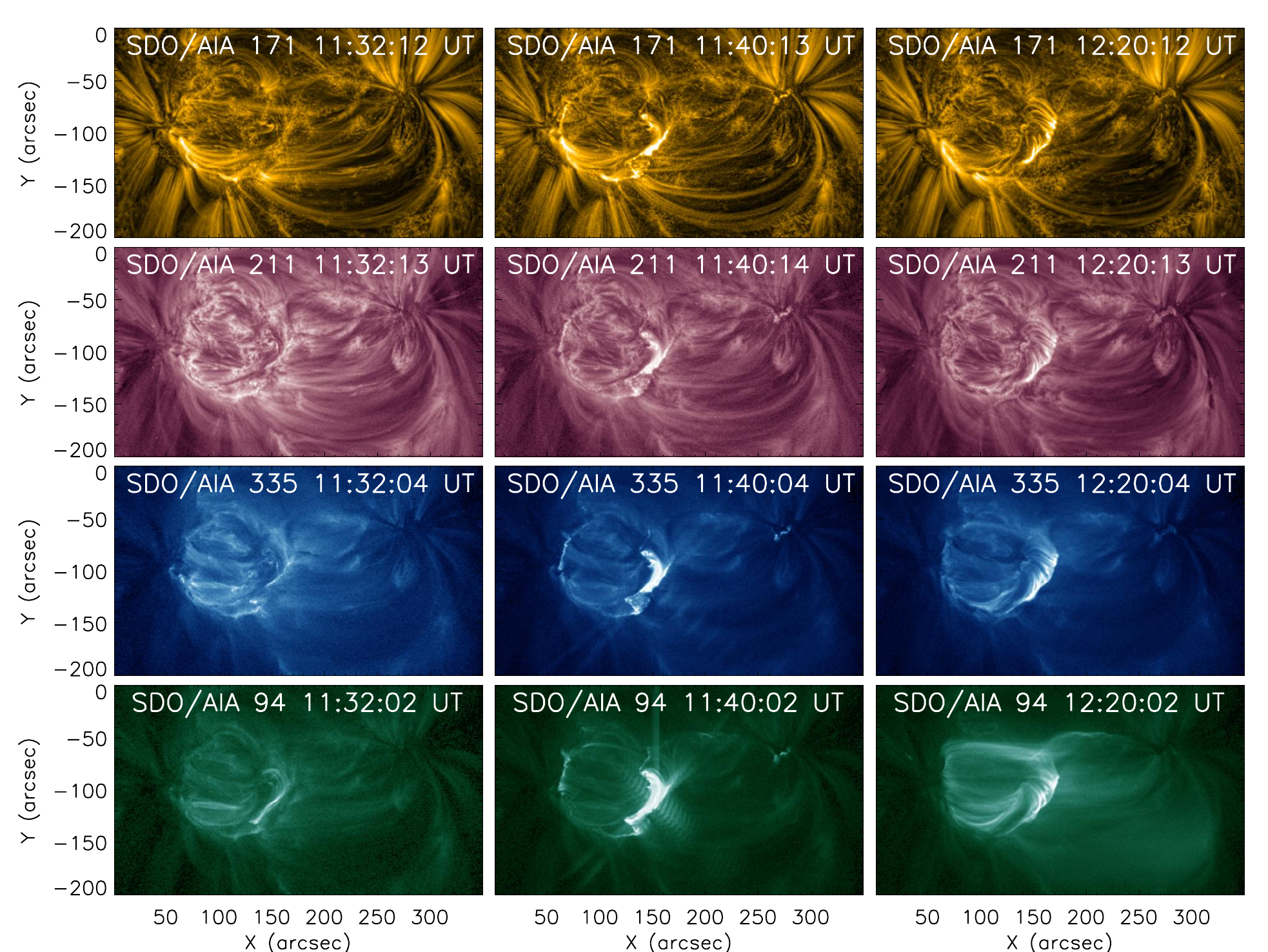}
\put(-460,295){\sf \wt{$\sim$0.6\,MK}}
\put(-460,205){\sf \wt{$\sim$2\,MK}}
\put(-460,115){\sf \wt{$\sim$2.5\,MK}}
\put(-460,25){\sf \wt{$\sim$6.3\,MK}}
}
\caption{
Time evolution of the coronal EUV emission in the course of the M2.1 flare. From left to right, snapshots of the pre-flare, impulsive and decay phase are shown. The emission at transition-region and quiet-corona temperatures (at 171~\AA) is shown in the top row. Emission from the active-region corona, at 211 and 335~\AA, is shown in the second and third row, respectively. The bottom row shows the emission from the flaring corona, at 94~\AA. The peak of the temperature response for each of the channels is shown at the bottom left of the first image for each row. In the Electronic Supplementary Material, a movie is attached to this figure. Movie 2 shows the evolution of the flare in co-temporal 304, 171, 193, 335, 94 and 131 \AA \ maps.
}
\label{multi_wavelength_sequence}
\end{figure*}
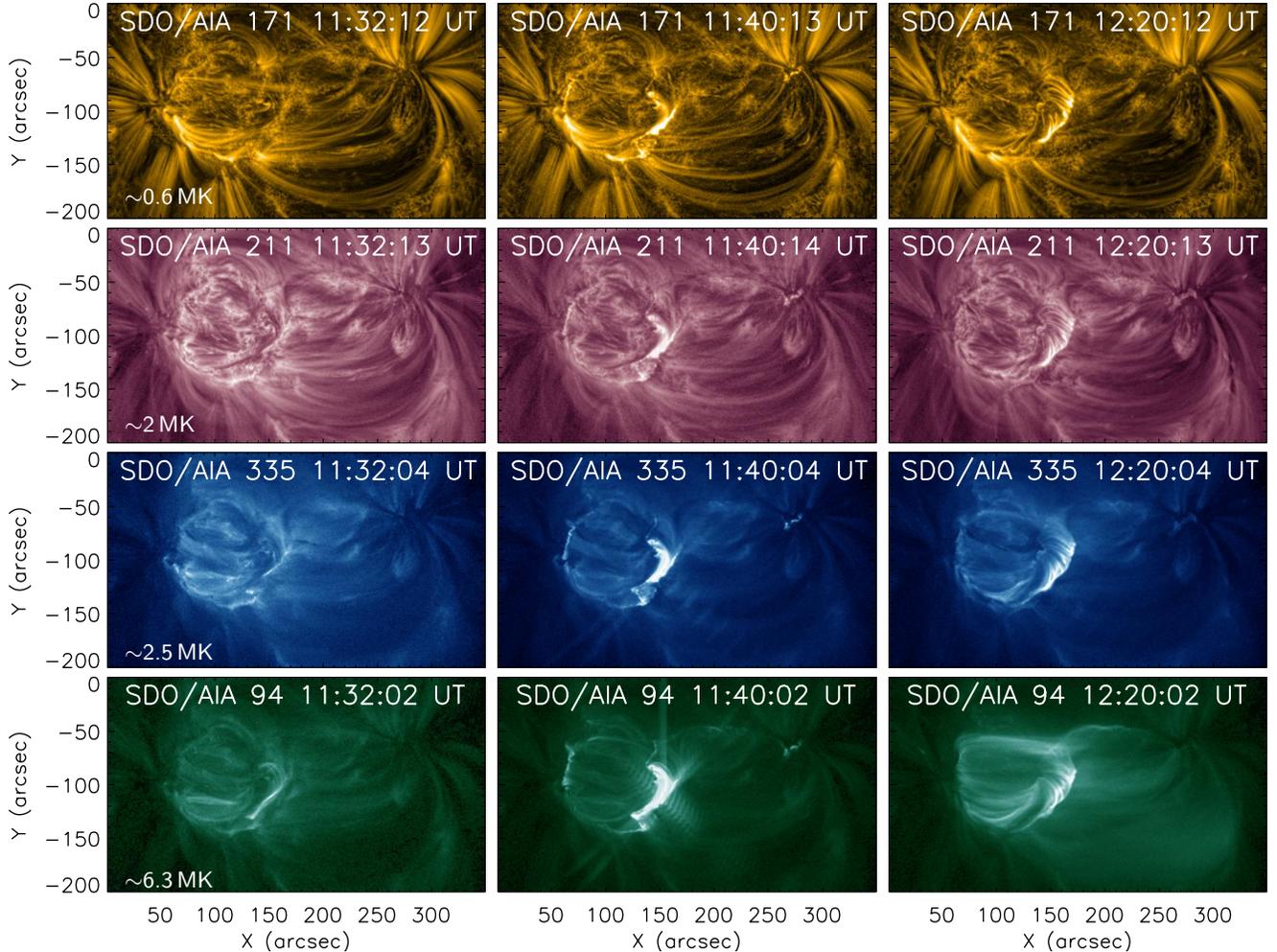

%%%%%%%%%%%%%%%%%%%%%%%%%%%%%%%%%%%%%%%%%%%%%%%%%%%%%%%%%%%%%%%%%%%%%%%%%%%%%%%%%%

The flare-induced changes to the plasma density and temperature in the flare loops, as observed at EUV wavelengths, are shown in Fig.~\ref{multi_wavelength_sequence}, where brightness maps characteristic for the pre-flare (left panels), early decay (middle panels) and late decay phase (right panels) are shown. During the impulsive phase, the plasma is heated and loops become visible in all EUV channels. Therefore, loops that were not visible during the pre-flare phase are visible during the early decay phase. During the late decay phase, bright arcades at the primary flare site are visible in all EUV channels, indicating that not only are they hotter than during the pre-flare phase and therefore visible in the channels with higher temperature response (\eg\ 335~\AA \ and 94~\AA) but also denser and therefore observed in the channel with lower temperature response (\eg\ AIA 171~\AA). Furthermore, during the decay phase, we see that the large-scale loops in the south-west of the AR fade at 171~\AA, 211~\AA \ and 335~\AA, and become brighter at AIA 94~\AA. This indicates that the plasma within these loops is hotter during the decay phase in comparison with the pre-flare phase. At 12:20:02~UT, all AIA channels clearly show a post-flare arcade, which connects the primary flare ribbons observed at UV wavelengths (compare Fig.~\ref{SDO_AIA_1600_SDO_AIA_304_LINEAR}). Furthermore, the plasma within the large-scale loops in the south-west of the AR, which apparently connect R1 and R3, are heated to high temperatures during the flare, this can best bee seen in AIA 94~\AA,. In the Electronic Supplementary Material, a movie is attached to Fig.~\ref{multi_wavelength_sequence}. Movie 2 shows the evolution of the flare at 304, 171, 193, 335, 94 and 131 \AA. The chromospheric ribbons emit strongly in 94 \AA\ during the early impulsive phase, and therefore possibly contributing strongly to the \goes\ SXR emission at that time \citep[\eg,][]{1994ApJ...422L..25H,2013ApJ...771..104F}.

%%%%%%%%%%%%%%%%%%%%%%%%%%%%%%%%%%%% FIGURE 4 %%%%%%%%%%%%%%%%%%%%%%%%%%%%%%%%%%%%
\begin{figure*} [ht]
\centerline{
\includegraphics[width=\textwidth]{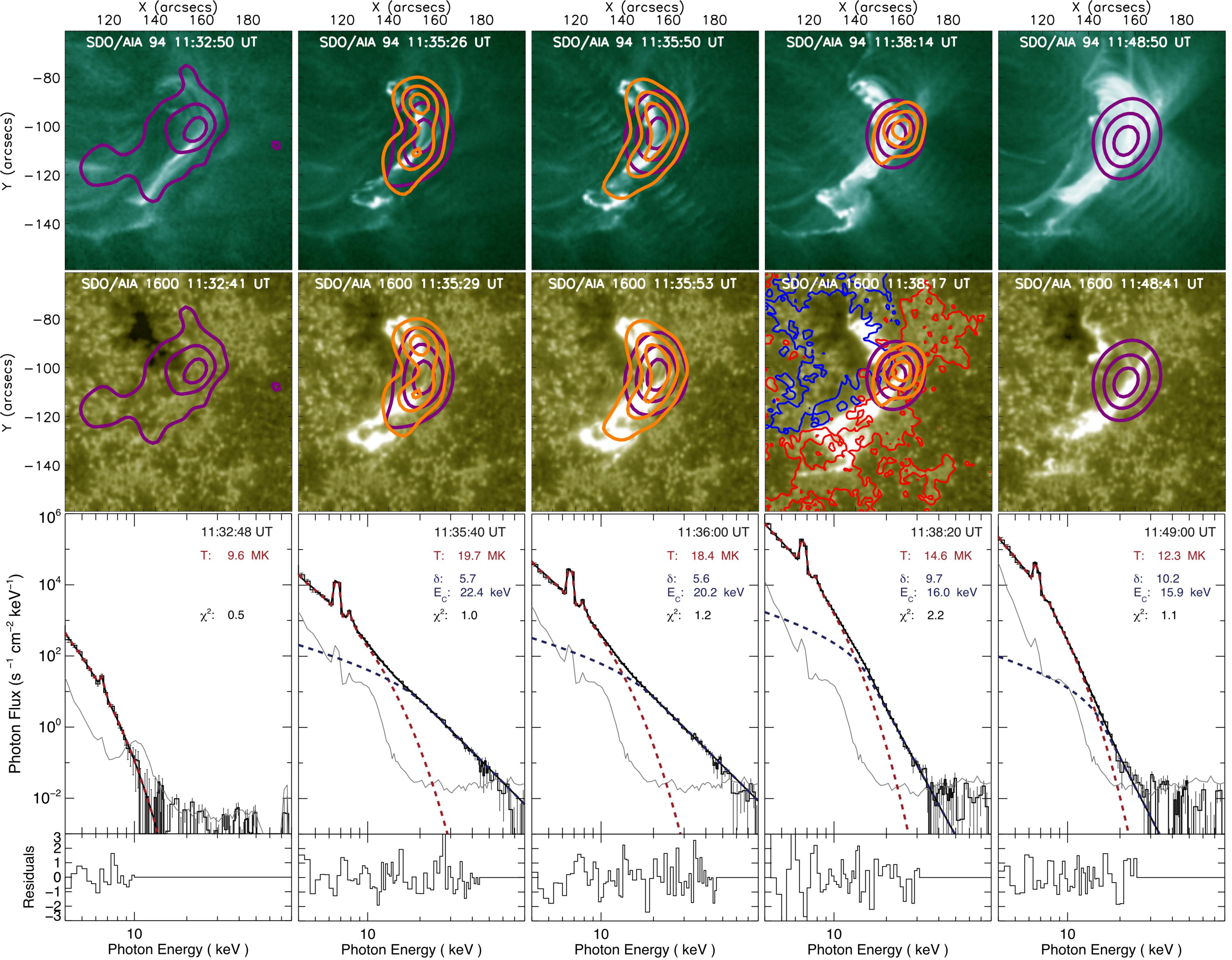}
\put(-445,405){\sf\color{black}(a)}
\put(-348,405){\sf\color{black}(b)}
\put(-251,405){\sf\color{black}(c)}
\put(-154,405){\sf\color{black}(d)}
\put(-57,405){\sf\color{black}(e)}
}
\caption{Time evolution of the flare-associated AIA EUV (at 94~\AA), UV (at 1600~\AA) and \rhessi\ X-ray emission contours (from top to bottom panels). From left to right, the pre-flare, impulsive (second--fourth column) and early decay phase (right column) is shown. At each time, contours are drawn at 45, 70, and 90\% of the maximum at thermal (6--12~keV; purple contours) and non-thermal (25--50~keV; orange contours) \rhessi\ X-ray energies. For comparison, the $B_{\rm z}=100$ G contour level of the positive (red) and negative (blue) photospheric LOS magnetic field magnitude is shown in one of the panels. The bottom panels show the \rhessi\ X-ray spectra (black solid lines) and fitting results for the isothermal component (red dashed lines) and the non-thermal component (blue dashed lines) for the corresponding flare phases. The background is represented by the gray solid line. The electron temperature, $T$, electron distribution index, $\delta$, and cutoff energy, $E_{C}$, are listed in the top right corner of each panel.
}
\label{rhessi}
\end{figure*}

%%%%%%%%%%%%%%%%%%%%%%%%%%%%%%%%%%%%%%%%%%%%%%%%%%%%%%%%%%%%%%%%%%%%%%%%%%%%%%%%%%

\rhessi\ X-ray spectra depicting the evolution of the flare-induced plasma heating and energized electrons for the times indicated by black arrows in the top panel of Fig.~\ref{SDO_AIA_1600_SDO_AIA_304_LINEAR}, are shown in the bottom panels of Fig.~\ref{rhessi}. The pre-flare phase (represented in Fig.~\ref{rhessi}a) is characterized by the absence of a non-thermal electron population, so the X-ray spectrum is best fitted by a purely isothermal component. At that time, the temperature of the emitting plasma is $\sim$9.6 MK. In the early impulsive phase
(Fig.~\ref{rhessi}b), the temperature increases to $\sim$19.7~MK, and the X-ray spectrum can no longer be fitted as purely isothermal. Instead, a power-law, non-thermal component appears. The hardest spectrum (Fig.~\ref{rhessi}c) occurred at the time of the main peak in the \rhessi\ 25--50~keV count rate (compare Fig.~\ref{SDO_AIA_1600_SDO_AIA_304_LINEAR}), with an electron power-law index of $\delta=5.6$, indicating significant non-thermal emission above $\sim$20~keV. During the decay phase (Fig.~\ref{rhessi}e), a non-thermal component is still present. 

\subsection{Spatial and temporal correspondence of UV and HXR emission}
     \label{S-Spatial and temporal correspondences of UV and HXR emission}

%%%%%%%%%%%%%%%%%%%%%%%%%%%%%%%%%%%% FIGURE 5 %%%%%%%%%%%%%%%%%%%%%%%%%%%%%%%%%%%%

\begin{figure}[ht!]
\centerline{
\includegraphics[width=0.5\textwidth,clip=]{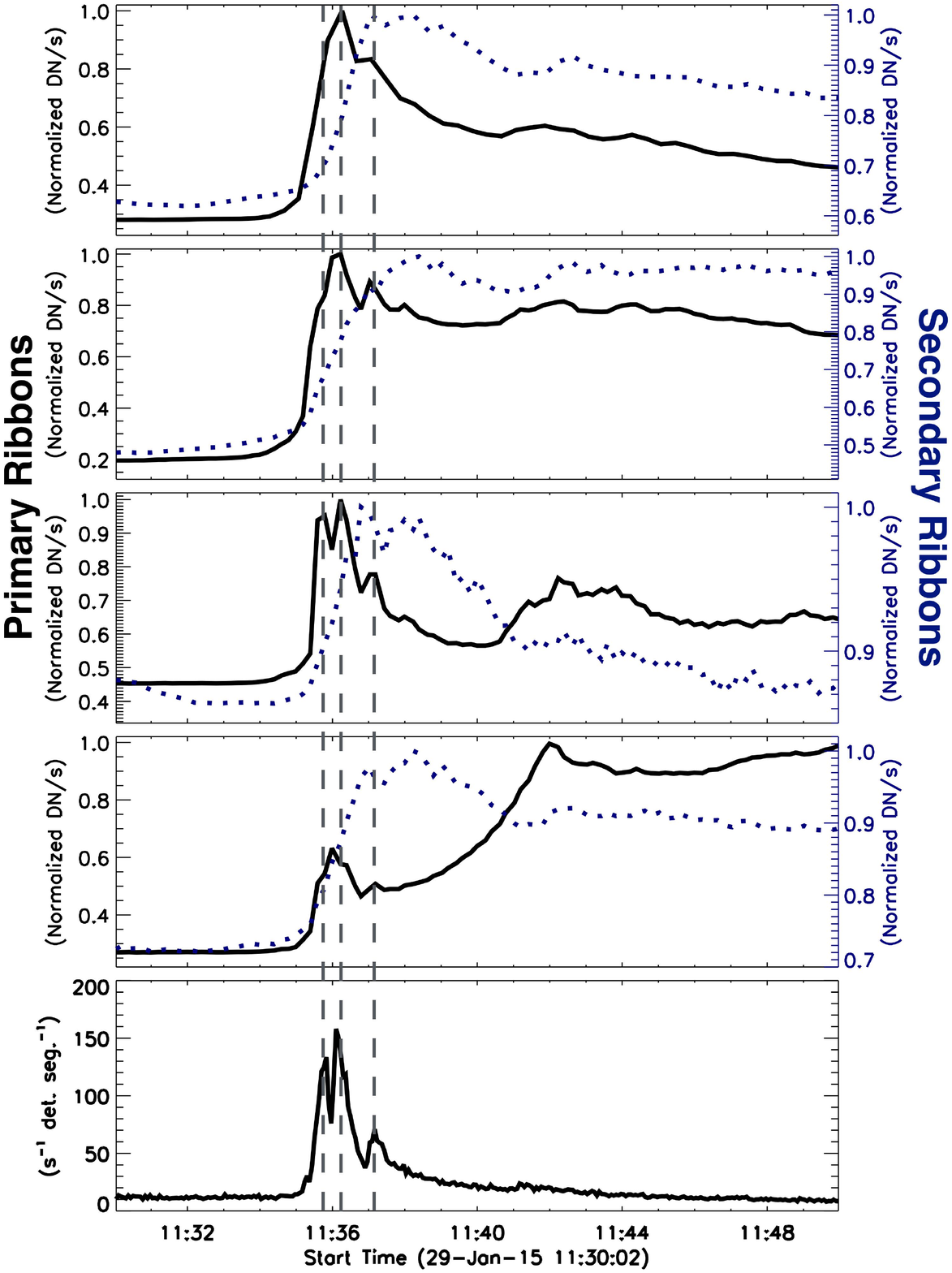}\\
\put(-222,338){\sf (a)}
\put(-222,273){\sf (b)}
\put(-222,208){\sf (c)}
\put(-222,143){\sf (d)}
\put(-222,78){\sf (e)}
}
\caption{
Total brightness of the primary (marked as ``R1'' and ``R2'' in Fig.~\ref{SDO_AIA_1600_SDO_AIA_304_LINEAR}) and secondary (marked as ``R3'' and ``R4'' in Fig.~\ref{SDO_AIA_1600_SDO_AIA_304_LINEAR}) flare ribbons (in black solid and dotted blue respectively), as a function of time (the selected regions for the lightcurves are shown in Fig.~\ref{continuum_LOS_AIA1600_as}c). From top to bottom, the total brightness in (a) 1600~\AA, (b) 304~\AA, (c) 171~\AA, and (d) 335~\AA\ is shown. Note that saturated and blooming pixels were excluded from the analysis. In (e), the \rhessi\ count rate in the 25--50~keV energy band is shown. The dashed lines indicate the time of the three \rhessi\ HXR peaks.
}
\label{lightcurves}
\end{figure}

%%%%%%%%%%%%%%%%%%%%%%%%%%%%%%%%%%%%%%%%%%%%%%%%%%%%%%%%%%%%%%%%%%%%%%%%%%%%%%%%%%  
 
As shown in Figure~\ref{SDO_AIA_1600_SDO_AIA_304_LINEAR} (top panel), three distinct HXR bursts occurred in the 25--50~keV energy band during the impulsive phase of the flare. An enhancement is also seen at even higher non-thermal energies (50--100~keV), though much less pronounced. The increase at low energies (6--12~keV and 12--25~keV), dominated by thermal emission from the hot flaring corona, corresponds to the  integral effect of the non-thermal emission, according to the so-called Neupert effect \citep[][]{1993SoPh..146..177D, 2002A&A...392..699V}.
 
In order to determine the location of thermal and non-thermal coronal X-ray sources, we construct \rhessi\ images of the entire AR. Fig.~\ref{rhessi} shows a sequence of images at chromospheric (1600~\AA) and coronal (94~\AA) temperatures, covering the region of primary ribbons, R1 and R2 (compare Fig.~\ref{continuum_LOS_AIA1600_as}c), together with the contours of the \rhessi\ X-ray sources. 

Prior to the flare onset, at 11:32:40~UT, a thermal X-ray source is present (purple contour, outlining the 6--12~keV emission). The emission stems from a highly sheared and/or twisted arcade of hot coronal loops, as seen in the 94~\AA \ image (Fig.~\ref{rhessi}a), which connects to locations in the low atmosphere where later the primary flare ribbons were observed (Fig.~\ref{rhessi}b). Corresponding pre-flare activity seems evident also from the enhanced level in the 6--25~keV \rhessi\ count rates (see Fig.~\ref{SDO_AIA_1600_SDO_AIA_304_LINEAR}). Between 11:35~UT and 11:38~UT, \ie\ during the impulsive phase, an extended non-thermal source appears (orange contours in Fig.~\ref{rhessi}b--\ref{rhessi}d). The strongest non-thermal emission is found in the form of two HXR kernels at the approaching ends of the flare ribbons (Fig.~\ref{rhessi}b), indicating that flare-accelerated electrons caused the observed ribbon emission via collision with the denser plasma at the chromospheric legs of the flaring loops.

Notably, no HXR sources were detected near the sites where the secondary ribbons (\eg\ R3 and R4) formed. This may be due to the dynamic range of \rhessi\ \citep[$\sim$ $10:1;$][]{2004ApJ...612..546S}, meaning that the imaging algorithm is not be able to accurately determine sources with an intensity below $\sim 10\%$ of the intensity of the brightest source at the same time. Since the most prominent (E)UV emission of the secondary ribbons (at $\sim11:38~UT$) is observed only one minute after the third HXR burst, we cannot rule out that the physical cause of the secondary ribbons is not non-thermal bremsstrahlung. In order to clarify the mechanism that caused the observed secondary flare ribbons, we study the relative timing of the ribbon-associated (E)UV and X-ray emission.

We select one region that contains the primary ribbons and two more regions containing each of the secondary ribbons (see red dashed boxes in Fig.~\ref{continuum_LOS_AIA1600_as}b) and calculate the total brightness (for the channels that did not undergo blooming), \ie, we integrate the intensity over the (core) flare region, as a function of time and for different temperature (wavelength) regimes. The total brightness for the primary ribbons, from chromospheric temperatures to hot flare plasma (black solid lines in Fig.~\ref{lightcurves}a--d), shows a close resemblance to the \rhessi\ 25--50~keV count rate during the early impulsive phase of the flare (compare Fig.~\ref{lightcurves}e). The distinct peaks in HXRs, are also evident in the integrated brightness curves, most pronounced at 171~\AA. Similar to the HXR emission, the brightness in the primary flare ribbon area quickly decays after the main HXR peak at $\sim$11:36~UT (within $\sim$1--2~minutes). This is consistent with the expected flare-related signatures in the flare model: flare-accelerated electrons penetrate the low atmosphere where they heat the chromospheric plasma via Coulomb collisions, which then expands and fills the post-flare loops with hot plasma (as seen at high temperatures). Furthermore, as seen for 335~\AA, an increase in brightness follows after the three HXR bursts reaching its maximum at the early decay phase (\ie, 11:42:00~UT) indicative of more plasma at high temperature after reconnection. This is easily observed in the movies provided in the on-line Supplementary Material (see Movie 2). The brightness at the location of the secondary ribbons (blue dotted lines in Fig.~\ref{lightcurves}a--d), reveals a first peak in the integrated (E)UV emission at the time of the last HXR peak (at $\sim$11:37~UT; compare Fig.~\ref{lightcurves}a--d to \ref{lightcurves}e). 
Importantly, a second peak in the total brightness at all wavelengths presented (more pronounced for AIA 304 and 335~\AA) is observed at $\sim$11:38:00~UT (\ie, $\sim$1~min after the last HXR burst), with a clearly longer decay time of $\gtrsim$5 minutes. This indicates a heating of the plasma around R3 and R4 due to a different process than Coulomb heating by electron beams, as indicated also by the absence of \rhessi\ HXR sources at those places, as discussed above. 

For completeness, we note that a type-III radio burst was recorded at 11:37~UT, corresponding in time with the third HXR peak registered by \rhessi, one minute before the secondary ribbons became most prominent. Careful inspection of the EUV image sequences, however, did not reveal obvious signatures of flare-related jet activity near the remote ribbon R3 (which would support the theory of magnetic reconnection at a coronal null point).  The causal link between the observed radio and HXR emission therefore remains elusive.

\subsection{Observation of early flare flow-like structures}
     \label{S-Observation of early flare flow-like structures}

%%%%%%%%%%%%%%%%%%%%%%%%%%%%%%%%%%%% FIGURE 6 %%%%%%%%%%%%%%%%%%%%%%%%%%%%%%%%%%%%

\begin{figure*}
\centerline{
\includegraphics[width=\textwidth,clip=]{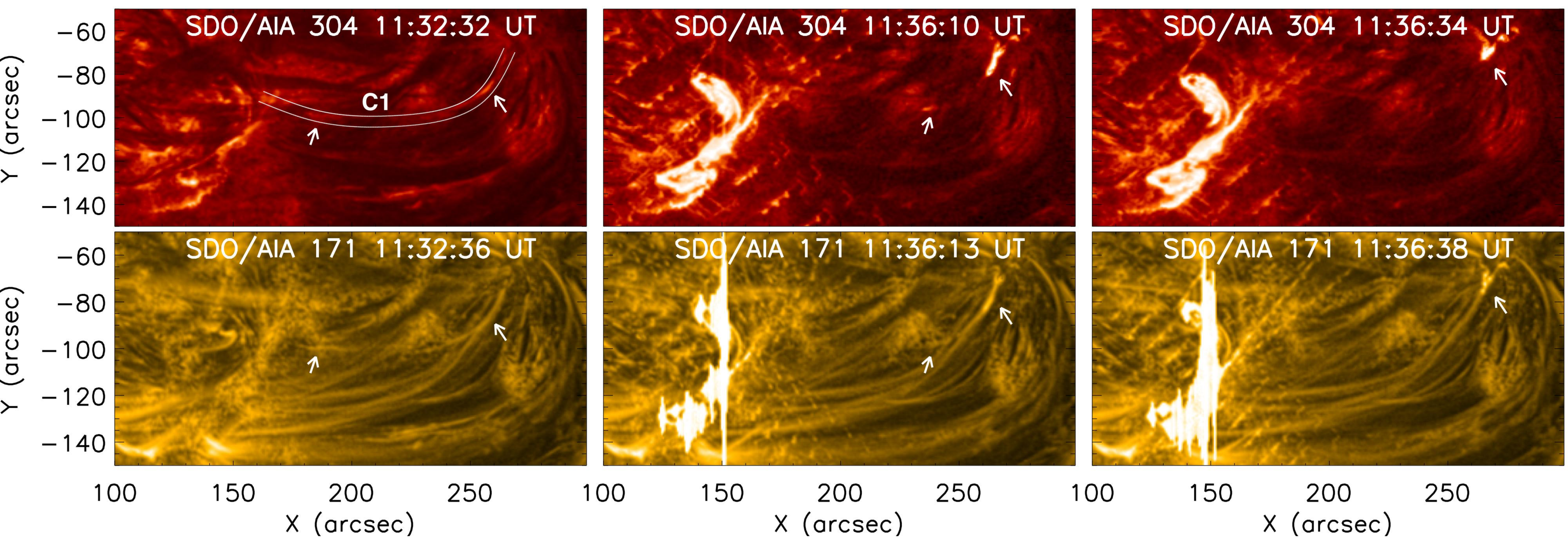}
} 
\caption{
Observed plasma motion between the negative-polarity primary (quasi-parallel) flare ribbon R1 and the remote secondary flare ribbon R3, during the early impulsive phase. Top and bottom panels show AIA 304 and 171~\AA\ images, respectively. White arrows indicate the position of individual plasma flows at the respective times. The two white lines outline the trajectory (C1) followed by the plasma flows. In the Electronic Supplementary Material, a movie is attached to this figure. Movie 3 shows the evolution of the flare in co-temporal AIA 304, 171, 193, 335, 94 and 131 \AA \ maps showing the motion of the flows of plasma.
} 
\label{blob_motion}
\end{figure*}

%%%%%%%%%%%%%%%%%%%%%%%%%%%%%%%%%%%% FIGURE 7 %%%%%%%%%%%%%%%%%%%%%%%%%%%%%%%%%%%%

\begin{figure*}
\centerline{
\includegraphics[width=\textwidth,clip=]{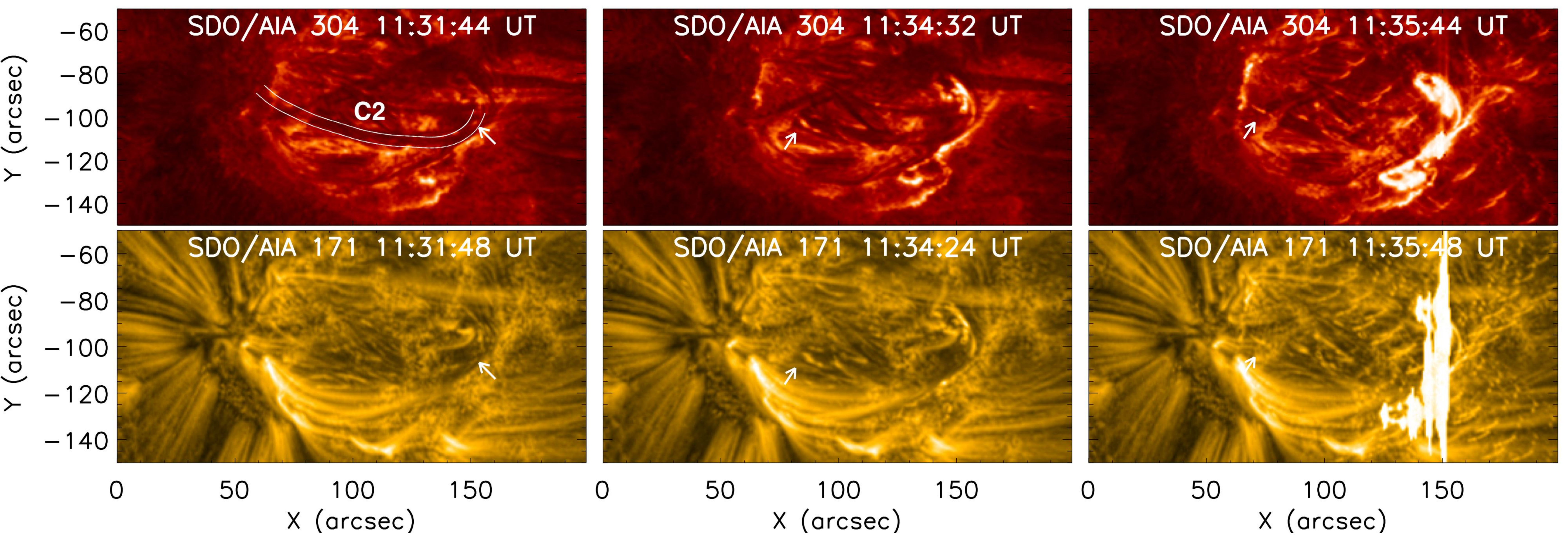} 
}
\caption{
Same as Fig.~\ref{blob_motion} but for a region containing R2 and R4. The two white lines outline the trajectory (C2) followed by the plasma flow. In the Electronic Supplementary Material, a movie is attached to this figure. Movie 4 shows the evolution of the flare in co-temporal AIA 304, 171, 193, 335, 94 and 131 \AA \ maps showing the motion of the flows of plasma.
}
\label{blob_motion2}
\end{figure*}

%%%%%%%%%%%%%%%%%%%%%%%%%%%%%%%%%%%%%%%%%%%%%%%%%%%%%%%%%%%%%%%%%%%%%%%%%%%%%%%%%%

Since no HXR sources were detected at the location of the secondary ribbons and the timing of the secondary flare ribbons is such that precipitating electrons seems to be an unlikely explanation for their generation, alternatives to the scenario involving flare-accelerated electrons are necessary.

During the impulsive phase, a number of flow-like structures traveling along loops are observed at EUV temperatures. They originate at the main flare site and travel towards the secondary ribbon sites. More precisely, flow-like structures that originate nearby R1 (the quasi-parallel ribbon located in the negative magnetic polarity; compare Fig.~\ref{SDO_AIA_1600_SDO_AIA_304_LINEAR}) terminate at the location where R3 (the remote flare ribbon) forms. Flow-like structures that originate nearby R2 (the quasi-parallel ribbon located in the positive magnetic polarity) terminate at the location where R4 (part of the quasi-circular flare ribbon) forms.

Figure~\ref{blob_motion} shows a sequence of EUV images revealing the motion of the aforementioned flow-like structures between the primary ribbons and R3 (indicated by white arrows), as seen at chromosphere and transition-region (at 304~\AA; upper panels) and coronal temperatures (at 171~\AA; lower panels). Figure~\ref{blob_motion2} shows the corresponding time sequence for the plasma structures between the primary ribbons and R4. In the Electronic Supplementary Material, two movies are attached to Figs.~\ref{blob_motion} and \ref{blob_motion2}. Movie 3 and 4 show the time evolution of the flare at 304, 171, 193, 335, 94 and 131 \AA, in which the flow-like structures are observed to travel towards the locations of the secondary ribbons, at times before these have formed. Interestingly, the observed structures are also detectable prior to the flare start and terminate temporally before but co-spatial with the secondary ribbons.

\begin{figure}[ht!]
\centerline{
\includegraphics[width=0.5\textwidth]{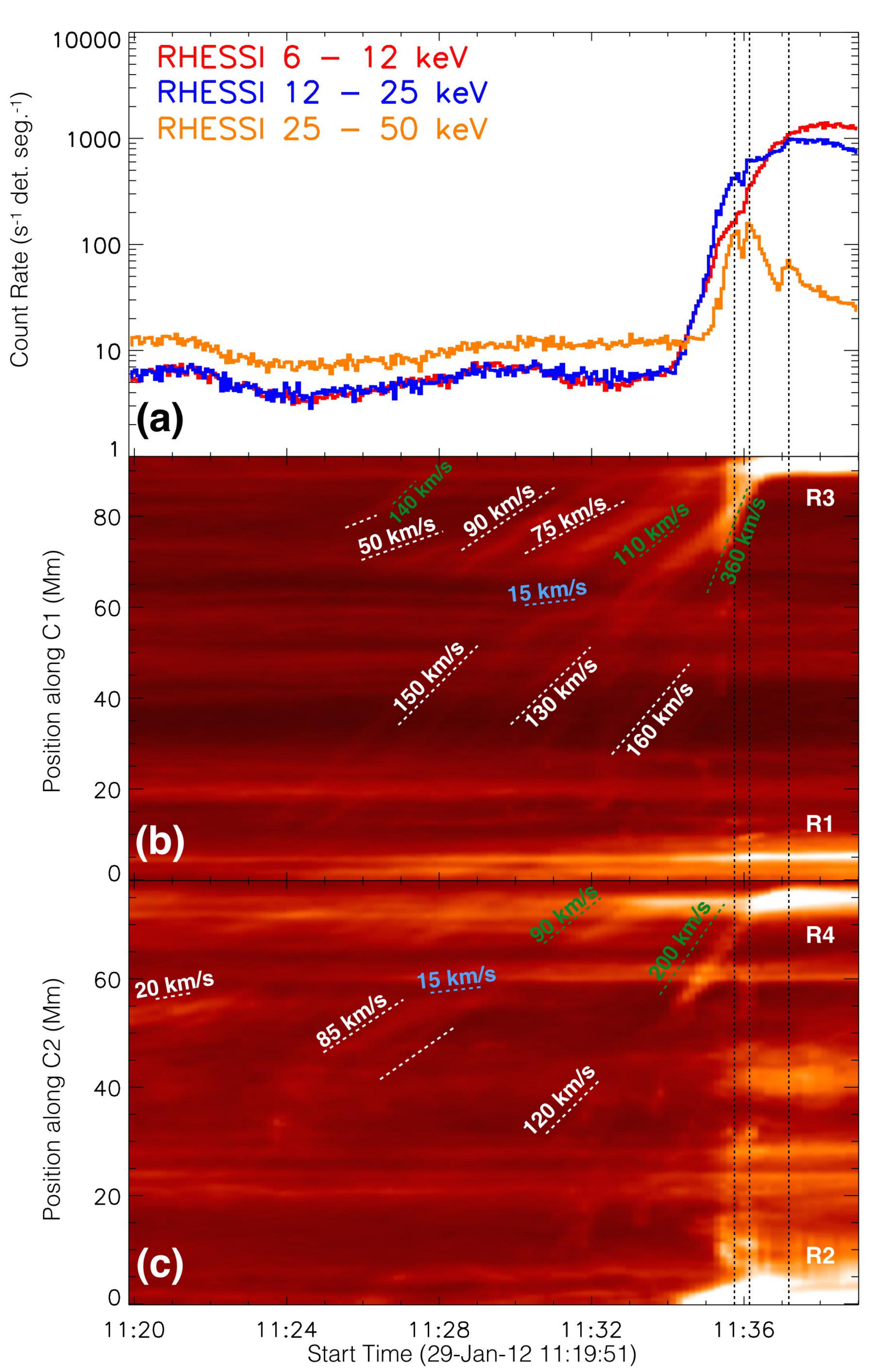}
}
\caption{
Time-distance profiles of the  AIA 304~\AA\ intensity along the selected paths C1 and C2 (compare Figures~\ref{blob_motion} and \ref{blob_motion2}), connecting (b) the primary ribbon R1 to the remote ribbon R3, and (c) connecting the primary ribbon R2 to the secondary ribbon R4. For comparison, the \rhessi\ 6--12~keV (red), 12--25 keV (blue) and 25--50~keV (orange) X-ray count rates are shown in the top panel.
}
\label{f:stack_plot_RHESSI_velocities_002}
\end{figure}

%%%%%%%%%%%%%%%%%%%%%%%%%%%%%%%%%%%%%%%%%%%%%%%%%%%%%%%%%%%%%%%%%%%%%%%%%%%%%%%%%%
In order to determine the speed of these structures we perform a time-distance analysis along the paths outlined in Figures~\ref{blob_motion} and \ref{blob_motion2} (\ie, C1 and C2). Fig.~\ref{f:stack_plot_RHESSI_velocities_002}b and \ref{f:stack_plot_RHESSI_velocities_002}c show stack plots of the intensity as observed in AIA 304~\AA, along the paths connecting R1 and R3 (C1), and R2 to R4 (C2), respectively. It can clearly be seen that the secondary (remote) ribbons are formed once the fastest structure ($\sim$360~km/s) arrives at the footpoints. Importantly, less-pronounced and less rapid structures extend back in time well into the pre-flare (early) phase of the flare and are clearly detectable as early as 11:23~UT (\ie\ $\sim$15 min before the onset). 

From the projected path of the moving structures, we find a velocity range of $\sim$15--360~km/s. These velocities represent lower limits since they were estimated based on 2D projections of true paths along 3D coronal loops. Assuming the inclination of these 3D loops and the solar surface to be $\sim$55$^{\circ}$ (based on a NLFF coronal magnetic field model; see Sect.~\ref{magnetic_field_analysis}), we are able to deduce a more realistic range of velocities of the flow-like structures. We find that the fastest EUV flow-like structure along C1 arrives with a speed of $\sim$630~km/s at R3.

Deceleration as well as acceleration of the structures (marked with blue and green dashed lines respectively in Fig.~\ref{f:stack_plot_RHESSI_velocities_002}a,b) was observed. Interestingly, the acceleration seems to occur right near their termination sites (suggesting gravitational acceleration).
 
%%%%%%%%%%%%%%%%%%%%%%%%%%%%%%%%%%%% FIGURE 8 %%%%%%%%%%%%%%%%%%%%%%%%%%%%%%%%%%%%

%%%%%%%%%%%%%%%%%%%%%%%%%%%%%%%%%%%%%%%%%%%%%%%%%%%%%%%%%%%%%%%%%%%%%%%%%%%%%%%%%%

%%%%%%%%%%%%%%%%%%%%%%%%%%%%%%%%%%%% FIGURE 9 %%%%%%%%%%%%%%%%%%%%%%%%%%%%%%%%%%%%

\begin{figure}[ht!]
\centerline{
\includegraphics[width=0.47\textwidth]{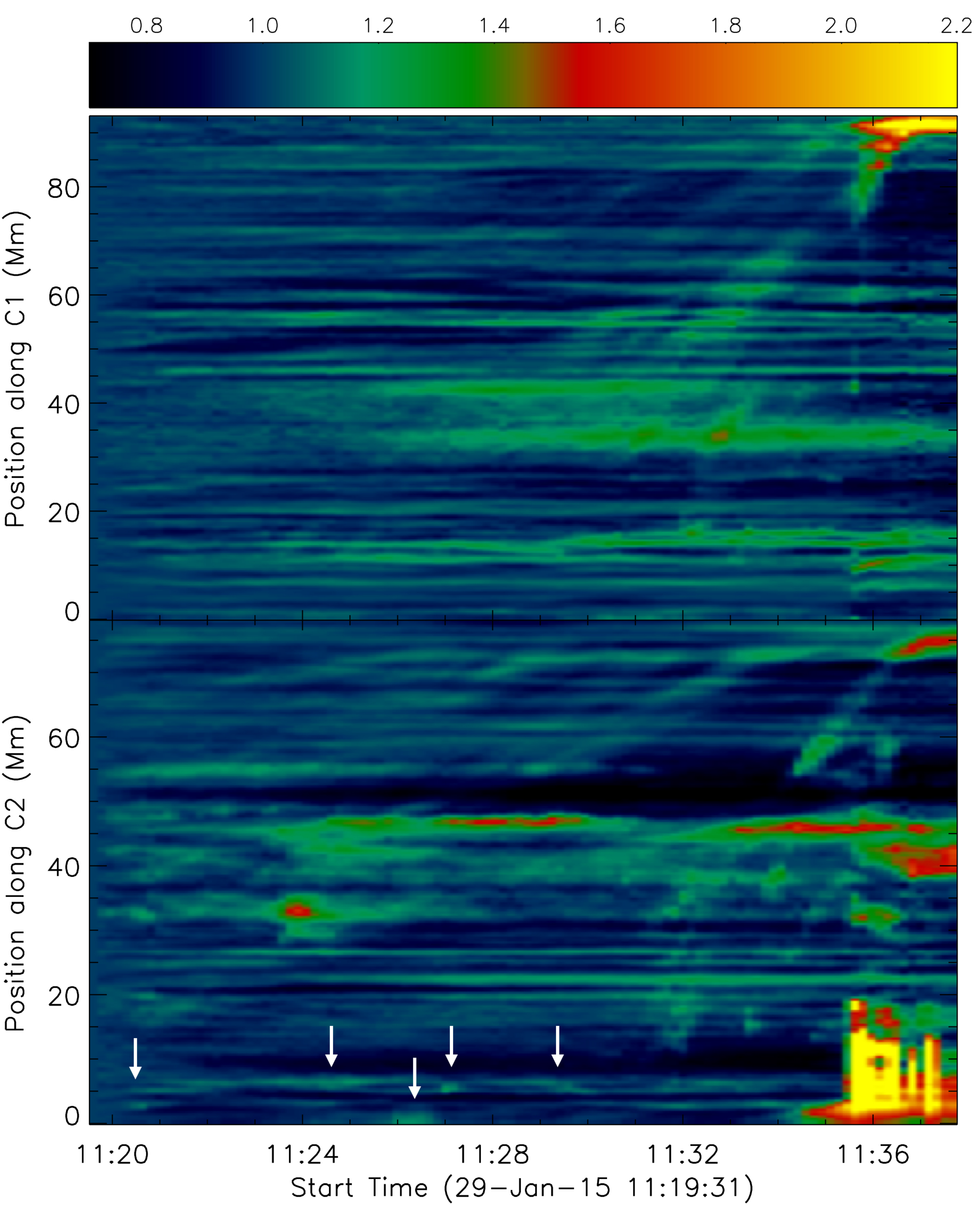}
}
\caption{
Base ratio time-distance profiles of the  AIA 171~\AA\ intensity along the selected paths C1 and C2 (compare Figures~\ref{blob_motion} and \ref{blob_motion2}), connecting (top) the primary ribbon R1 to the remote ribbon R3, and (bottom) connecting the primary ribbon R2 to the secondary ribbon R4. The white arrows indicate transient brightenings that occurred at the primary flare site.
}
\label{f:stack_plot_RHESSI_helvetica_percent_4_171}
\end{figure}

%%%%%%%%%%%%%%%%%%%%%%%%%%%%%%%%%%%%%%%%%%%%%%%%%%%%%%%%%%%%%%%%%%%%%%%%%%%%%%%%%%

In order to study these structures more in depth, we performed base ratio time-distance plots (base image at 11:19:31~UT) to see the changes with respect to the initial intensities. For this analysis we chose AIA 171~\AA\ (not in 304 \AA\ where they are best observed) because it is optically thin and the intensities observed for this channels are the intensities integrated along the line of sight, being able to picture the changes along the selected paths. Fig.~\ref{f:stack_plot_RHESSI_helvetica_percent_4_171} shows the base ratio time-distance plots for both C1 and C2 for AIA 171~\AA. We observe that the traces indicating the motion of the structures exhibit an increase in intensity of $\sim$30--40\%. The intensity increase at the location of the secondary ribbons is of more than 100\% for R1 and of $\sim$60--70\% for R2.

Another interesting observational finding is that at the starting point of C2 ($[x,y]=[152\asecs,-90\asecs$]) transient brightenings that occur during the early flare are observed in AIA 131 \AA\ (see Movie 4). This location, as seen in Fig.~\ref{continuum_LOS_AIA1600_as}b corresponds to a small region ($2\asecs \times 2\asecs$) of negative polarity. The signature of those early flare transient brightenings are associated with an intensity increase of $\sim$20--30\% (marked with white arrows) suggesting heating events during the early pre-flare phase.

\subsection{Physical nature of the EUV flow-like structures}
     \label{DEM_analysis}

%%%%%%%%%%%%%%%%%%%%%%%%%%%%%%%%%%%% FIGURE 10 %%%%%%%%%%%%%%%%%%%%%%%%%%%%%%%%%%%%

\begin{figure}
\centerline{
\includegraphics[width=0.5\textwidth]{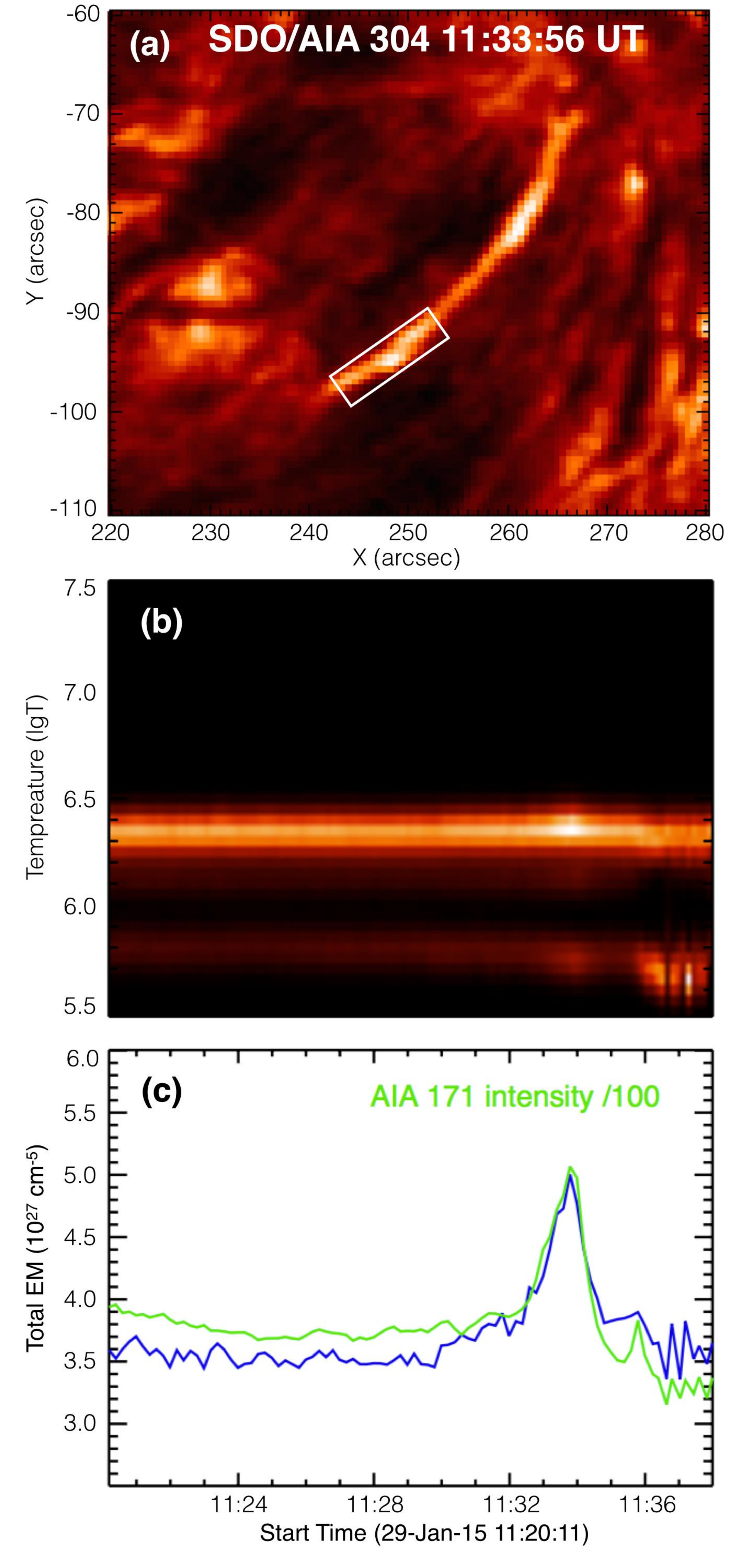}
}
\caption{
(a) AIA 304 image at the time where the EUV-emitting structures were best visible for this channel, the white rectangle represents the region considered for the DEM, (b) Time evolution of total EM in the region of interest, (c) The time evolution of the sum of EM(T) over the whole range of temperatures (in blue), compared with the light curve of AIA 171 intensities (in green). 
}
\label{f:DEM5}
\end{figure}

%%%%%%%%%%%%%%%%%%%%%%%%%%%%%%%%%%%%%%%%%%%%%%%%%%%%%%%%%%%%%%%%%%%%%%%%%%%%%%%%%%

In order to determine the physical nature of the observed flow-like features, we perform a DEM analysis. To do this, we calculated the Emission Measure of the most prominent observed structure (white rectangle in Fig.~\ref{f:DEM5}a) covering an area of $\sim$12\asecs\ in length and $\sim$3.6\asecs\ in width. This DEM analysis covers the temperature range $\log~T=5.5-7.5$ and suggests that the plasma contained in the flow-like feature is at $\lesssim~3MK$ ($log~T\sim6.4$; see Fig.\ref{f:DEM5}b). During the flare, the average EM increases by $\approx 1.5\times10^{27}cm^{-5}$ (see Fig.\ref{f:DEM5}c). Assuming that the depth of the structure along the line of sight is the same as the width, the average number density is \(n= \sqrt[]{1.5 \times 10^{27}~cm^{-5}/depth}= 2.4 \times 10^{9}~cm^{-3}\). The ratio of the density of the structure to the density of the background can be estimated by
\[\frac{\rho_{structure}}{\rho_{background}}=\sqrt[]{\frac{EM_{peak}}{EM_{background}}}=1.2\]

Since the background plasma has a larger depth along the LOS this means that the density increased by a factor of at least 20\%. We estimated the total mass of the structure to be of $\sim$$4.0 \times 10^{8}~kg$, assuming that 75\% of the plasma is hydrogen and the rest is helium. Therefore, at a speed of 630~km/s (when arriving at the location where R3 forms), the structure has kinetic energy of about $8 \times 10^{26}$~ergs. This is the energy for the fastest EUV-emitting structure, although we also found several (slower) structures in the time-distance plot that would add up to this energy. Finally, the peak thermal energy \citep{2005JGRA..11011103E} in the ribbon R3 based on DEM analysis is of the order of $10^{27}$~ergs (assuming the ribbon has a depth of 2\asecs).

In the following, we try to explain the observed features based on the inherent coronal and underlying photospheric magnetic field structure and evolution.

\subsection{Flare-associated coronal magnetic field structure}
     \label{magnetic_field_analysis}
     
%%%%%%%%%%%%%%%%%%%%%%%%%%%%%%%%%%%% FIGURE 11 %%%%%%%%%%%%%%%%%%%%%%%%%%%%%%%%%%%

\begin{figure*}
\centerline{
\includegraphics[width=\textwidth]{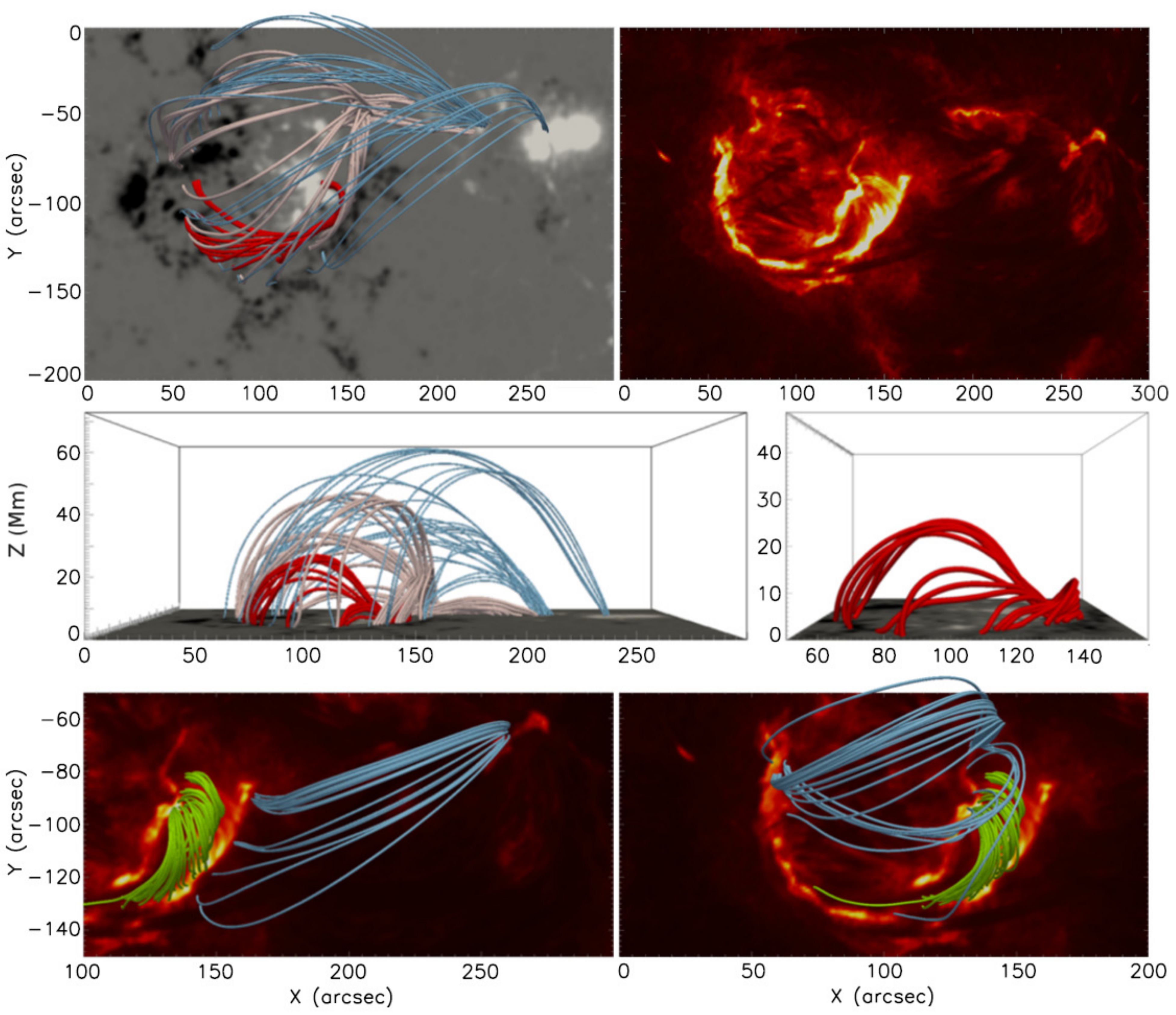}\\
\put(-470,410){\sf\color{white}\Large\bf(a)}
\put(-235,410){\sf\color{white}\Large\bf(b)}
\put(-470,240){\sf\Large\bf(c)}
\put(-165,240){\sf\Large\bf(d)}
\put(-470,120){\sf\color{white}\Large\bf(e)}
\put(-235,120){\sf\color{white}\Large\bf(f)}
} 
\caption{
(a) Coronal magnetic field connectivity prior to the M2.1 flare (around 10:24~UT). Sample field lines are shown, calculated from low-atmosphere regions that were later populated by flare ribbons. The grayscale background resembles the vertical component of the NLFF lower boundary magnetic field, scaled to $\pm1000$~G. Black/white indicates negative/positive polarity. Blue lines outline the field connecting the remote ribbon R3, located in the leading positive-polarity sunspot, and the horseshoe-shaped negative-polarity region. Pink lines represent the field connecting the E-W aligned intermediate ribbon, located around $(x,y)=(200\asecs,-50\asecs)$, and the negative polarity region, exhibiting the shape of a fan-like dome. The red lines outline a low-lying magnetic flux rope, located beneath the south-western part of the overlying fan-like field. (b) Post-flare AIA 304~\AA\ emission. (c) Side view (along the positive solar-$y$ direction) of the coronal magnetic field configuration. (d) Zoomed-in view of the flux rope underlying the coronal fan. Same viewing direction as in (c). (e)/(f) Post-flare magnetic field (around 12:12~UT) outlining the magnetic connectivity between the primary (quasi-parallel) ribbons R1 and R2 (green lines) and the remote ribbons R3 and R4 (blue lines). Units are arc-seconds from Sun center.
}
\label{f:mf_modeling}
\end{figure*}

%%%%%%%%%%%%%%%%%%%%%%%%%%%%%%%%%%%%%%%%%%%%%%%%%%%%%%%%%%%%%%%%%%%%%%%%%%%%%%%%%%

Rapid emergence of positive-polarity magnetic flux in the eastern part of the AR led to the evolution of a positive-polarity region, surrounded by a quasi-circular (horseshoe-shaped) rim of negative polarity (see Fig.~\ref{continuum_LOS_AIA1600_as}). The associated model coronal magnetic field exhibited a prominent fan-like shape. In Fig.~\ref{f:mf_modeling}a, we show sample field lines, calculated from positions where flare ribbons were observed at later times (for comparison see the post-flare AIA 304~\AA\ emission in Fig.~\ref{f:mf_modeling}b). 

The pre-flare dome-like coronal magnetic field in the eastern part of the AR (see pink lines in Fig.~\ref{f:mf_modeling}a and \ref{f:mf_modeling}c, forming a fan-like field) coincides with the orientation of the observed coronal loops (compare, \eg, lower panels of Fig.~\ref{blob_motion2}). The magnetic connections to the western part of the AR, where the leading positive-polarity sunspot was located, merge with the outer fan-like field to the north and south of the AR, and also bridge the entire configuration to close at the eastern part of the horseshoe-shaped negative polarity region (see blue lines in Fig.~\ref{f:mf_modeling}a and \ref{f:mf_modeling}c). Preliminary analysis of the underlying magnetic field topology suggests the presence of a low-lying null point around $(x,y)$=(160\asecs,-50\asecs) (courtesy of F. Zuccarello). A more detailed topological analysis of the flare-associated coronal field is outside of the scope of the present paper, and will be presented in a forthcoming study. The existence of the null point, however, is important for the subsequent discussion.

The emergence of positive flux caused its western parts to be swept towards the neighboring parts of the horseshoe-shaped negative-polarity region (around $(x,y)=(150\asecs,-100\asecs)$). Presumably as a consequence of flux cancellation at the corresponding part of the quasi-circular PIL, a low-lying magnetic flux rope developed as a result of thus induced tether-cutting \citep[\eg,][]{1989ApJ...343..971V}. The flux rope is recovered in the pre-flare NLFF model (see red lines in Fig.~\ref{f:mf_modeling}a, c and c) and located underneath the south-west of the fan-like field. It spatially coincides with a dark filament channel observed prior to the flare (compare the pre-flare AIA 211~\AA\ image shown in Fig.~\ref{multi_wavelength_sequence}).

As discussed above, the reconstructed NLFF field recovers prominent features observed in coronal images, such as the fan-like coronal loop configuration and a flux rope associated with an observed filament channel. Similarly, the sheared post-flare arcade, as observed at EUV wavelengths (see last column of Fig.~\ref{multi_wavelength_sequence}), is recovered in the form of a field arcade connecting the primary flare ribbons (\ie, R1 and R2) in a post-flare NLFF model (green lines in Fig.~\ref{f:mf_modeling}e and f). These spatial correspondences underline the quality of our magnetic field model and motivates us to study the magnetic connectivities of the primary and secondary flare ribbons. Magnetic field that stems from around the location of the remote ribbon R3 (located within the positive-polarity sunspot) connects to locations nearby R1 (Fig.~\ref{f:mf_modeling}e). Note that field lines were calculated from footpoints located around $(x,y)=(265\asecs,-70\asecs)$, \ie, the western end of the path used to trace plasma flows towards R3 (compare upper left panel of Fig.~\ref{blob_motion} and  Fig.~\ref{f:stack_plot_RHESSI_velocities_002}b). From the sample field lines shown in Fig.~\ref{f:mf_modeling}f, it is also evident that the observed plasma flows traveling towards the secondary ribbon R4 stem from the vicinity of the primary ribbon R2. The field lines were traced from the eastern end of the path used to trace the corresponding plasma flows (see upper left panels of Fig.~\ref{blob_motion} and  Fig.~\ref{blob_motion2}). These findings suggest that the observed secondary (remote) ribbon signatures were not a direct consequence of electrons accelerated at the coronal null point impinging on the low solar atmosphere. 

Thus an alternative explanation is required for the ribbon and flow features observed. As outlined above, the pre-flare NLFF coronal magnetic field model does not reveal a direct connectivity between the position of the remote ribbon R3 and the presumed location of the fan-associated null point, which one would expect if the remote ribbon R3 were caused by reconnection at the coronal null. We suspect that, as the null-point associated dome grows in time fueled by the ongoing emergence of positive magnetic flux in its center, the outermost parts of the fan dome are driven towards the opposite-polarity surrounding magnetic field. Magnetic reconnection on small-scales with the ambient field may have caused the observed flows towards R3, and finally lead to the delayed remote ribbon brightening there.

%%%%%%%%%%%%%%%%%%%%%%%%%%%%%%%%%%%%%%%%%%%%%%%%%%%%%%%%%%%%%%%%%%%%%%%%%%%%%%%%%%

\section{Discussion} 
      \label{S-Discussion} 

The event under study (SOL2015-01-29T11:42) exhibited a complex ribbon configuration, in which two quasi-parallel primary ribbons form in the early impulsive phase, followed by the formation of two secondary ribbons (see Fig.~\ref{SDO_AIA_1600_SDO_AIA_304_LINEAR}). EUV observations revealed that one of the quasi-parallel primary ribbons and one of the secondary ribbons are segments of an extended quasi-circular ribbon. The other secondary ribbon exhibited an elongated shape and was located at a remote location. The observed signatures may be interpreted as consequences of magnetic reconnection of a coronal null point \citep[\eg,][]{2013ApJ...778..139S,2015ApJ...812L..19L,2016A&A...591A.141J, 2016ApJ...832...65Z}. Detailed study of the spatial and temporal organization of the associated emission, DEM analysis and NLFF modeling of the underlying magnetic field, however, suggest an alternative explanation for the physical process by which the secondary ribbons form.

A distinct SXR source was observed above the primary ribbons (Fig.~\ref{rhessi}), and was present throughout the entire flare. Furthermore, during the impulsive phase, non-thermal emission was observed at the same location in the form of two HXR kernels, on top of the approaching ends of the flare ribbons, indicative of accelerated electrons impinging on the chromosphere (Fig.~\ref{rhessi}b--\ref{rhessi}d). No \rhessi\ emission was detected at the sites where the secondary ribbons formed. This could either be due to \rhessi's limited dynamic range, or because the chromospheric brightenings at those locations were not produced by electron beams but another physical mechanism. 

The HXR and (E)UV emission associated to the primary (quasi-parallel) ribbons shows the expected behavior: a clear time-correspondence is seen, in response to the burst-like HXR emission (black solid lines in Fig.~\ref{lightcurves}). This is not the case, however, for the (E)UV and X-ray emission from the secondary flare ribbons: the peak (E)UV intensity lags the HXR bursts up to $\gtrsim2$ minutes (blue dashed lines in Fig.~\ref{lightcurves}). This is different from what was previously reported in the literature \citep[\eg,][]{2009ApJ...700..559M,2012A&A...547A..52R,2015ApJ...806..171Y}. In these works, the emission from the secondary/remote ribbons was also nearly co-temporal with HXR peaks, suggesting a common driver in the form of electrons that were accelerated at a coronal null point configuration subject to magnetic reconnection. Moreover, in contrast to \cite{2012A&A...547A..52R}, we did not find a thermal branch of emission outlining a newly established magnetic field linkage to the remote ribbon.

Our event revealed a number of multi-thermal EUV-emitting structures that originate at the primary flare site and travel along loops towards the secondary flare sites at speeds of up to $\sim$630~km/s. The arrival times of these multi-thermal structures at the secondary flare sites are co-temporal with the enhanced (E)UV enhanced ribbon emission. Furthermore, time-distance plots showed that these features extend back in time well into the pre-flare phase, detectable as early as $\sim$15 minutes before the impulsive flare onset. The fact that the observed structures originated at the primary flare site could indicate a pressure imbalance due to sudden heating events that occurred during the early flare phase.

Movie 4 shows that during the pre-flare and early flare phase, a number of transient brightenings occur in the vicinity of the primary flare site. At the starting point of C2 ($[x,y]=[155\asecs,-90\asecs$]) we observe transient brightenings visible in AIA 131 \AA. Transient brightenings at the primary flare site are observed in 171 \AA\ as intensity increases of $\sim$20--30\% during the early flare (marked with white arrows in Fig.~\ref{f:stack_plot_RHESSI_helvetica_percent_4_171}).

Since the transients are observed as emission enhancements, it is possible that the EUV-emitting structures are either a traveling pulse \citep[\eg, ][]{2013A&A...558A..76R} that could have been generated due to the null point \citep[][]{2017A&A...602A..43S} or plasma flows. \cite{2002A&A...387L..13D} presented a statistical study of 38 events in which longitudinal oscillations (slow magnetosonic waves) in large coronal loops were observed. The study revealed that these disturbances only propagate upwards at an almost constant speed of about 65-165 km/s (no acceleration or deceleration was observed) with an emission intensity perturbation always below 10\%. This variation in intensity was observed for 171 \AA, and for only 2 out of the 38 events were these disturbances observed in 195 \AA. Finally, these disturbances were found to be more prominent at the beginning, and their intensity decreased as they moved upwards.

In comparison to the results of \cite{2002A&A...387L..13D} (see Fig.~\ref{f:stack_plot_RHESSI_velocities_002}) we found several different characteristics in the present study: 1. The speeds of the traveling EUV structures reach up to $\sim$630~km/s; 2.~Both deceleration and acceleration (see Fig.~\ref{f:stack_plot_RHESSI_velocities_002}b and c) was observed, probably due to the effect of gravity since these accelerations are registered in the vicinity of the termination site close to the solar surface; 3.~An intensity increase of $\sim$30--40\% (see Fig.~\ref{f:stack_plot_RHESSI_helvetica_percent_4_171}), corresponding to more than thrice the maximum intensity increase observed in \cite{2002A&A...387L..13D}; 4.~Density increase of the EUV-emitting structures of at least $\sim$20\%, such large increases would point toward large-amplitude or shock waves; 5.~Downward motion of these features was observed; 6.~Enhancement in the intensity as they move downwards; 7.~They were observed in all EUV channels, indicative of their multi-thermal nature. These differences suggest that the moving structures do not represent EUV disturbances and leaves flows of plasma as the most plausible scenario.

DEM analysis (Fig.~\ref{f:DEM5}) was performed in order to compare the kinetic energy of the fastest plasma flow arriving at the secondary flare site and the thermal energy of the secondary ribbon at that location. The thermal energy for R3 was found to be of the order of $10^{27}$~ergs, and the kinetic energy of the last and fastest flow of plasma arriving at the same location was found to be of $\sim 8.0 \times 10^{26}$~ergs, this consistency supports plasma flows as the physical cause of the secondary ribbons.

Such early flows of plasma have recently been observed and reported in \cite{2015ApJ...812L..19L}. In their study of the \goes-class X1.0 flare on 2014 March 29, they observed flows of plasma from the reconnection region flowing towards the chromosphere. Although not discussed explicitly in their study, one can see a response in the ribbon-integrated lightcurve \citep[RC1 in][]{2015ApJ...812L..19L} at the time where plasma flows collide with the chromosphere during the early phase ($\sim$6 min before the main HXR peak). This response is not very pronounced due to the large region considered to perform the lightcurve. Furthermore, no SXR or HXR signatures were detected at this location.

Another scenario of enhanced ribbon-like emission at EUV wavelengths has been discussed by \eg, \cite{2001SoPh..204...69F}, where they explain that such emissions may not be entirely attributable to the bombardment of the low solar atmosphere with flare-accelerated electrons. Instead, some might develop due to heat conduction, progressing from a heated flare loop top towards the low atmosphere. These processes may be recognized based on the time evolution of the associated (E)UV emission, in comparison to that of the flare-associated X-rays. In the first scenario (initiation due to electron bombardment), a close time-correlation (with maybe a delay of the order of seconds) is expected as a consequence of the impulsive start and end of the reconnection-driven particle acceleration process. In the second scenario (conduction-driven from the flare loop top), due to the fact that the heat front needs time to travel to the footpoints, a time-dependence between the EUV signature at the loop-top and the HXR peak at the footpoints of the order of minutes may be detected. However, in our event, the previous results suggest that heating by means of conduction fronts is an unlikely explanation for the enhancements at the secondary flare sites.

Furthermore, if the secondary ribbons were caused by the same reconnection process as the primary ones, then it is difficult to explain why the flows terminating at the site of the secondary ribbons and causing their maximum brightness, are already detectable before the flare onset (see Fig.~\ref{f:stack_plot_RHESSI_velocities_002}). This highlights the importance of physical processes prior to the flare, \ie, during the pre-flare (early) phase, and the necessity of related detailed studies in the future.

Finally, NLFF modeling (Fig.~\ref{f:mf_modeling}) did not reveal direct magnetic connectivity between the presumed coronal null point and the secondary ribbons. Instead, magnetic fields emerging from the secondary ribbons terminated in the photospheric periphery of the fan-like coronal magnetic field, closely resembling the apparent paths of the observed plasma flows.

The findings presented in this paper suggest that the generation of the secondary ribbons did not occur due to non-thermal particles accelerated by magnetic reconnection. Instead, the most probable alternative scenario, is that the moving plasma compressed the chromospheric material at the secondary flare sites, dissipating its kinetic energy, and therefore causing the enhanced emission at these locations.

\section{Conclusion} 
      \label{S-Conclusion} 

The event under study exhibits a fan-spine coronal magnetic configuration, in which two quasi-parallel primary ribbons form in the early impulsive phase, followed by the formation of two secondary ribbons at remote locations. The results indicate that the enhanced emission at the secondary flare sites was generated by a different physical mechanism to the standard explanation of electron beams colliding with the chromosphere, as a consequence of magnetic reconnection.

We propose an alternative physical interpretation, where the heating at the low-atmosphere footpoints of newly reconnected fields during the early flare phase, associated with the primary ribbons, produces an overpressure, that thermally drives flows of plasma along neighboring coronal loops of differing magnetic connectivity. Once the moving plasma arrives at the secondary ribbon sites, it compresses the chromospheric material, dissipating its kinetic energy, thus causing the enhanced (E)UV emissions.

For the primary flare site we observe a group of reconnected sheared arcades connecting the two primary ribbons during the decay phase. \rhessi\ thermal emission (during the early, impulsive and decay phases) and non-thermal emission (during the impulsive phase) coming from the primary flare site was detected, and strong time correlation was found between the total (E)UV brightness and the \rhessi\ 25--50~keV. Additionally, we find direct magnetic connectivity between a low-lying null-point and the primary flare site, as evidenced by the nonlinear force-free model. These results indicate that the primary ribbons were generated by thick-target bremsstrahlung, in the chromosphere, as a result of magnetic reconnection.

However, there are several pieces of evidence that suggest a different mechanism for the secondary flare site. Firstly, neither thermal nor non-thermal X-ray emission at the secondary flare sites was detected, as evidenced by the absence of \rhessi\ sources, and also, the maximum (E)UV emission of the secondary ribbons occurs 1 minute after the last 25--50~keV peak registered by \rhessi. A number of multi-thermal plasma flows were generated during the early flare at the primary flare site that were observed to travel along loops towards the secondary flare sites. A direct correlation between the formation of the secondary ribbons and the arrival of these plasma flows at the secondary sites was found. An analysis of intensity and density increase of these plasma flows, with respect to background emission, was found to be of more than thrice what was previously reported in the literature. Additionally, no connectivity was found between the low-lying null point and one of the secondary flare sites. Finally, a DEM analysis shows a very close relationship between the thermal energy of one of the secondary ribbons and the kinetic energy of the fastest plasma flow arriving at that location. These observations can be explained by our proposed mechanism of heating due to compression. This scenario poses challenges to multiple-ribbon flare models and leaves room for different interpretations of secondary ribbons to that of magnetic reconnection.

\acknowledgments
\small

We thank the anonymous referee for careful consideration of this manuscript and helpful comments. JKT thanks G. Valori and F. Zuccarello for helpful discussions. This study was supported by the Austrian Science Fund (FWF): P25383-N27, P27292-N20, V195-N16, and by the \"{O}sterreichischer Austauschdienst (OeAD), the Slovak Research and Development Agency (SRDA): SK 01/2016,
SK-AT-2016-0002, the Scientific Grand Agency: VEGA 2/0004/16, the Thousand Young Talents Plan (a sub-program of the ”1000 Talent Plan”), and the Joint Research Fund in Astronomy (U1631242) under cooperative agreement between the National Natural Science Foundation of China (NSFC) and Chinese Academy of Sciences (CAS). \sdo\ is a mission for NASA's Living With a Star (LWS) Program. \sdo\ data are courtesy of the NASA/\sdo\ and HMI science team. \rhessi\ is a NASA Small Explorer Mission. \goes\ is a joint effort of NASA and the National Oceanic and Atmospheric Administration (NOAA).

%%%%%%%%%%%%%%%%%%%%%%%%%%%%%%%%%%%% BIBLIOGRAPHY %%%%%%%%%%%%%%%%%%%%%%%%%%%%%%%%

\bibliography{bibliography}   

\end{document}